\documentclass{sig-alternate}
\usepackage{mathptmx} 

\usepackage{fancyhdr}
\usepackage[normalem]{ulem}
\usepackage[hyphens]{url}
\usepackage[sort,nocompress]{cite}
\usepackage[final]{microtype}
\usepackage{flushend}

\usepackage{wrapfig}
\newcommand{\ignore}[1]{}
\usepackage{color}
\usepackage{soul}
\usepackage[htt]{hyphenat}

\usepackage[noend]{algpseudocode}
\usepackage{algorithm}
\usepackage{amsbsy}
\usepackage{amssymb}
\usepackage{tabularx}
\usepackage{subfig}
\usepackage{pifont}

\newcommand{\smallcapital}{\fontsize{9pt}{10pt}\selectfont}

\usepackage{balance}
\usepackage{multirow}
\usepackage{verbatim}

\usepackage[subtle]{savetrees}
\fancypagestyle{firstpage}{
  \fancyhf{}
}

\vspace{-0.1in}
\title{CuttleSys: Data-Driven Resource Management for Interactive Applications on Reconfigurable Multicores} 

\author{Neeraj Kulkarni, Gonzalo Gonzalez-Pumariega, Amulya Khurana, Christine Shoemaker,\\ Christina Delimitrou, and David Albonesi\\\normalsize Cornell University}

\begin{document}
\maketitle
\thispagestyle{firstpage}
\pagestyle{plain}


\begin{abstract}
Multi-tenancy for latency-critical applications leads to resource interference and unpredictable performance.
Core reconfiguration opens up more opportunities for colocation, as it allows the hardware to adjust to
the dynamic performance and power needs of a specific mix of co-scheduled applications.
However, reconfigurability also
introduces challenges, as even for a small number of reconfigurable cores,
exploring the design space becomes more time- and resource-demanding.

We present CuttleSys, a runtime for reconfigurable multicores that leverages
scalable and lightweight data mining to quickly identify suitable core and cache 
configurations for a set of co-scheduled applications. 
The runtime combines collaborative filtering to infer the behavior of each job
on every core and cache configuration, with Dynamically Dimensioned Search to efficiently 
explore the configuration space. 
We evaluate CuttleSys on multicores with tens of reconfigurable cores
and show up to 2.46$\times$ and 1.55$\times$ performance improvements compared to core-level gating and oracle-like asymmetric multicores respectively, under stringent power constraints.
\vspace{-0.08in}
\end{abstract}
\section{Introduction}

Cost efficiency in datacenters is adversely affected by low resource utilization~\cite{BarrosoBook,Delimitrou13,Delimitrou14,Delimitrou13b,Delimitrou13d,Delimitrou14b,Delimitrou15,Delimitrou16,Delimitrou17,Mars13a,Mars13b,Lo14,Lo15,Chen19,Delimitrou15,GoogleTrace}. 
Server utilization can be improved through multi-tenancy which, however, 
is especially challenging for latency-critical applications, such as websearch, social networks, and ML inference, since 
it can lead to interference in shared resources (cores, cache, memory bandwidth, network bandwidth, power, etc.), 
and unpredictable performance. Prior work has proposed techniques to avoid interference by disallowing colocation 
of contending workloads~\cite{Mars13a,Mars13b,Delimitrou13,Delimitrou14,Delimitrou16,Chen19}, 
or techniques to eliminate interference altogether, 
by leveraging hardware and software resource isolation mechanisms~\cite{Vantage11,Kasture14,kasture15,Chen19,Lo14,Lo15,Romero18,Hsu18,Pliant}. 

In multi-tenant systems with latency-critical applications, fine-grained resource allocation allows 
assigning just enough resources to co-scheduled applications to meet the QoS, which in turn, improves resource efficiency by allowing more applications to be co-scheduled.
However, prior work is limited 
to traditional servers where cores cannot be reconfigured to enable fine-grained performance and power adjustments. Core reconfiguration~\cite{flicker,chrysso,DCS}
opens up more opportunities for colocation, as it allows the hardware to adjust to the dynamic needs of a specific mix of co-scheduled applications. 

DVFS, which is widely used in systems today, is another solution to allow fine-grained performance and power adjustments in cores. 
However, the movement towards processors with razor-thin voltage margins and the
increase in leakage power consumption limits the effectiveness of DVFS in future systems~\cite{Skylake,CoffeeLake,Meisner11,pcpg,meisner09, meisner3}. 
Reconfigurable cores~\cite{flicker,chrysso,DCS} operate by dynamically power gating core components. Since they reduce both active and leakage power, 
they can be effective in reducing power consumption in technologies 
where voltage scaling ranges are limited. 
Datacenters also suffer from poor energy proportionality~\cite{Lo14,Meisner11}, 
stemming from the high idle power of processors as technology shrinks. 
Reconfigurable cores with their ability to reduce idle power, 
also offer a solution to make cloud servers more energy proportional. 



We propose to leverage reconfigurable cores to enable co-scheduling of latency-critical and batch applications. 
For scenarios that involve colocation of latency-sensitive and batch applications, this means 
satisfying the strict quality of service (QoS) requirements of interactive services, and maximizing the throughput of the 
batch applications, while always remaining under the allowed power budget assigned to the server either 
by the chip-wide power budget, or by a global power manager~\cite{Lo14} running datacenter-wide. 
Prior work~\cite{flicker} addresses reconfiguration exclusively for batch applications, 
and leads to QoS violations and unpredictable performance for latency-critical services. 
Additionally, Flicker~\cite{flicker} does not handle interference in the shared memory hierarchy. 
On the other hand, fine-tuning architectural parameters also increases the space of 
allocations a resource manager must traverse to identify suitable resource configurations for an application. 
As the number of cores and configuration parameters increase, efficiently 
exploring this space becomes computationally prohibitive. This is even more challenging given 
that decisions must be online, as applications and power budgets change. 


We design CuttleSys, an online resource manager that combines scalable machine learning 
to determine the performance and power of each application across all possible core and cache 
reconfigurations, with fast design space exploration to effectively navigate the large configuration space 
and arrive at a high-performing solution.
First, the system leverages collaborative filtering, namely PQ-reconstruction with Stochastic Gradient Descent (SGD) to infer the performance 
(tail latency for latency-critical and throughput for batch applications) 
and power consumption of an application across core and cache configurations without the overhead of exhaustive profiling. 
Second, it leverages a new, parallel Dynamically Dimensioned Search (DDS) algorithm  
to efficiently find a per-job near-optimal configuration that satisfies QoS for latency-sensitive workloads, 
and maximizes the throughput for batch jobs, under a given power budget. 
Both techniques keep overheads low, a couple milliseconds, allowing CuttleSys to reevaluate 
its decisions frequently to adjust to changes in application behavior. 

We make the following contributions: 
\vspace{-0.05in}
\begin{itemize}  \setlength\itemsep{-0.1em}
	\item We demonstrate the potential of reconfigurable cores for cloud servers when running latency-critical applications by characterizing five representative interactive cloud services (Section~\ref{sec:characterization}).
	\item We present CuttleSys, an online resource manager that efficiently navigates the large design space and determines suitable core and cache configurations (Section~\ref{sec:prob}).   
  \item We evaluate CuttleSys on 32-core simulated systems with mixes of latency-sensitive~\cite{tailbench} 
and batch applications~\cite{spec}. We show that at near-saturation load and across different power caps, 
CuttleSys achieves 2.46$\times$ higher throughput than core-level gating and 1.55$\times$ higher than 
an oracle-like asymmetric multicore, while always satisfying QoS for the latency-sensitive applications. 
We also show that CuttleSys effectively adapts to changes in input load and power budgets online (Section~\ref{sec:results}). 
\end{itemize}
\vspace{-0.12in}

\section{Related Work}


\vspace{-0.08in}
\subsection{Power Management}
\vspace{-0.08in}

\subsubsection{Dynamic Voltage-Frequency Scaling}

Dynamic Voltage-Frequency Scaling (DVFS) allows dynamically changing a processor's voltage and frequency, and is widely used in modern multicores. 

\noindent{\textbf{Batch Workloads}}:
Isci \emph{et al.}~\cite{Isci} propose maxBIPS, an 
algorithm that selects DVFS modes for each core that maximize 
throughput under a power budget. Sharkey \emph{et al.}~\cite{Sharkey} extend this work 
by exploring both DVFS and fetch toggling, as well as design tradeoffs such as local versus
global management. Bergamaschi \emph{et al.}~\cite{Bergamaschi:ASPDAC2008} further extend maxBIPS, 
and compare its discrete implementation to continuous power modes. 
Chen \emph{et al.} \cite{cmp_smt:John} propose co-ordinated predictive hill climbing to control distribution of power among cores,
and intra-core resources like IQ, ROB and register files among SMT threads.
{Papadimitriou \emph{et al.} \mbox{\cite{voltage-guardbands}} explore safe Vmin for different applications by exposing pessimistic guardbands 
and determining the best voltage, frequency, and core allocation at runtime. }

Apart from open-loop solutions, there are also multiple feedback-based controllers~\cite{feedback1,feedback2,feedback3,Lo14, twig}.  
Wang \emph{et al.}~\cite{feedback1} use Model Predictive Control to maintain the power of a CMP below the 
budget by controlling the DVFS states, while Bartolini \emph{et al.}~\cite{feedback3} propose a distributed solution allocating one 
MPC-based controller to each core. Ma \emph{et al.}~\cite{feedback2} propose a hierarchical solution for many-core architectures that 
divides the problem by allocating frequency budgets to smaller groups of cores. Intel also supports fine-grained power control through the 
RAPL~\cite{RAPL} interface that allows software to set a power limit, which the hardware meets by scaling voltage/frequency. 

\vspace{0.03in}
\noindent{\textbf{Latency Sensitive Workloads}}: 
Lo \emph{et al.}~\cite{Lo14} propose 
a feedback-based controller that reduces power consumption in server clusters, while meeting 
the QoS (Quality of Service) requirements of latency-critical services by adjusting the server power limits using RAPL.
{Nishtala \emph{et al.} \mbox{\cite{twig}} use Reinforcement Learning to find the best 
core allocations and frequency settings for latency-critical jobs to save energy while meeting QoS. }
Kasture \emph{et al.}~\cite{kasture15} propose Rubik, 
a fine-grained DVFS scheme for latency-sensitive workloads and RubikColoc, a scheme to co-schedule 
batch and latency-critical workloads. Adrenaline~\cite{adrenaline} applies DVFS 
at a per-query granularity, using application-level information to speed up long queries. 
Meisner \emph{et al.}~\cite{Meisner11} explore the efficacy of active and idle low-power 
modes for latency-critical applications to save power under QoS, and showed that 
active power modes (DVFS) provide good power-performance trade-offs but cannot achieve 
energy proportionality by themselves. 
Motivated by their conclusion, our work explores 
fine-grained power management techniques that reduce idle power along with active power.

The movement towards processors with razor-thin voltage margins 
limits the effectiveness of DVFS as technology scaling slows down. 
A viable and widely-implemented alternative to DVFS is core-level gating (C states), discussed in the next section. 
Reconfigurable cores enable gating at an even finer granularity allowing further gains over traditional core-level gating. 
Similar to how core-level gating is used along-side DVFS in modern processors, our technique can augment DVFS 
by increasing the energy gains for frequency regions where DVFS is not effective~\cite{DCS, Skylake, CoffeeLake}. 

\vspace{-0.08in}
\subsubsection{Core-Level Gating} 

Core-level gating powers off individual cores by placing them in a separate domain~\cite{SandyBridge,Ivybridge,Skylake,Kumar:ISSCC2009}, 
and has become necessary to reduce power consumption beyond DVFS. 
Intel CPUs since Skylake~\cite{Skylake,CoffeeLake}   
support Duty Cycling Control (DCC), which rapidly cycles between on $(C0)$ and off $(C6)$ states for each core at the granularity of tens of microseconds. 
A few of the proposals to use core-level gating to maximize performance under a power budget are described below.
 
\vspace{0.03in}
\noindent{\textbf{Batch Workloads}}:
Intel processors~\cite{Skylake,CoffeeLake} implement 
core-level gating only during idle core times using auto-demotion. 
Ma \emph{et al.}~\cite{PGCapping} and Huazhe \emph{et al.}~\cite{feedback4} 
integrate core-level gating with DVFS, and propose a controller-based algorithm 
that employs power gating at coarse granularity, and DVFS at fine granularity. 
Arora \emph{et al.}~\cite{feedback_tullsen} develop a linear prediction algorithm 
for $C6$ for CPU-GPU benchmarks. Pothukuchi \emph{et al.}~\cite{MIMO} use 
MIMO theory, while Rahmani \emph{et al.}~\cite{SPECTR} use Supervisory Control Theory 
to dynamically tune architectural parameters to meet performance and power goals. 
These feedback-based controllers become overly expensive as the decision space expands, 
taking a prohibitive time to converge. 

\vspace{0.03in}
\noindent{\textbf{Latency Sensitive Workloads}}: 
Leverich \emph{et al.}~\cite{pcpg} propose per-core power-gating to dynamically turn cores on/off 
based on utilization and QoS. 
PowerNap~\cite{meisner09} and DreamWeaver~\cite{meisner3} 
coordinate deep CPU sleep states to minimize idle power. 
However, Kanev \emph{et al.}~\cite{Kanev} show that deep CPU sleep states, 
owing to their long wakeup latencies, can also impact tail latency, as latency-sensitive 
applications have short idle periods. We use core-level gating in this work as a baseline 
for cores that host batch workloads to meet the power budget.

\vspace{-0.08in}
\subsection{Asymmetric Multicores}

Asymmetric multicores improve performance and power by assigning resources to applications based on their dynamic requirements~\cite{hetero1,hetero2,hetero3,hetero4,hetero5,hetero6,hetero7, hetero8}. 

\noindent{\textbf{Batch Workloads}}:
PIE~\cite{Emer:PIE} schedules applications in heterogeneous multicores by estimating
the performance of an application on out-of-order cores, while running on an in-order core and vice-versa.
Liu \emph{et al.}~\cite{Liu:2013} propose a dynamic thread-mapping approach, \textit{maximization-then-swapping},
to maximize performance in power-constrained heterogeneous multicores. However, this relies on application profiling, 
which can become impractical in large-scale multicores. 
\begin{figure*}[tbh]
	\centering
	\begin{tabular}{ccccc}
		\includegraphics[scale=0.166, bb=0 0 30 30,viewport= 60 0 480 390]{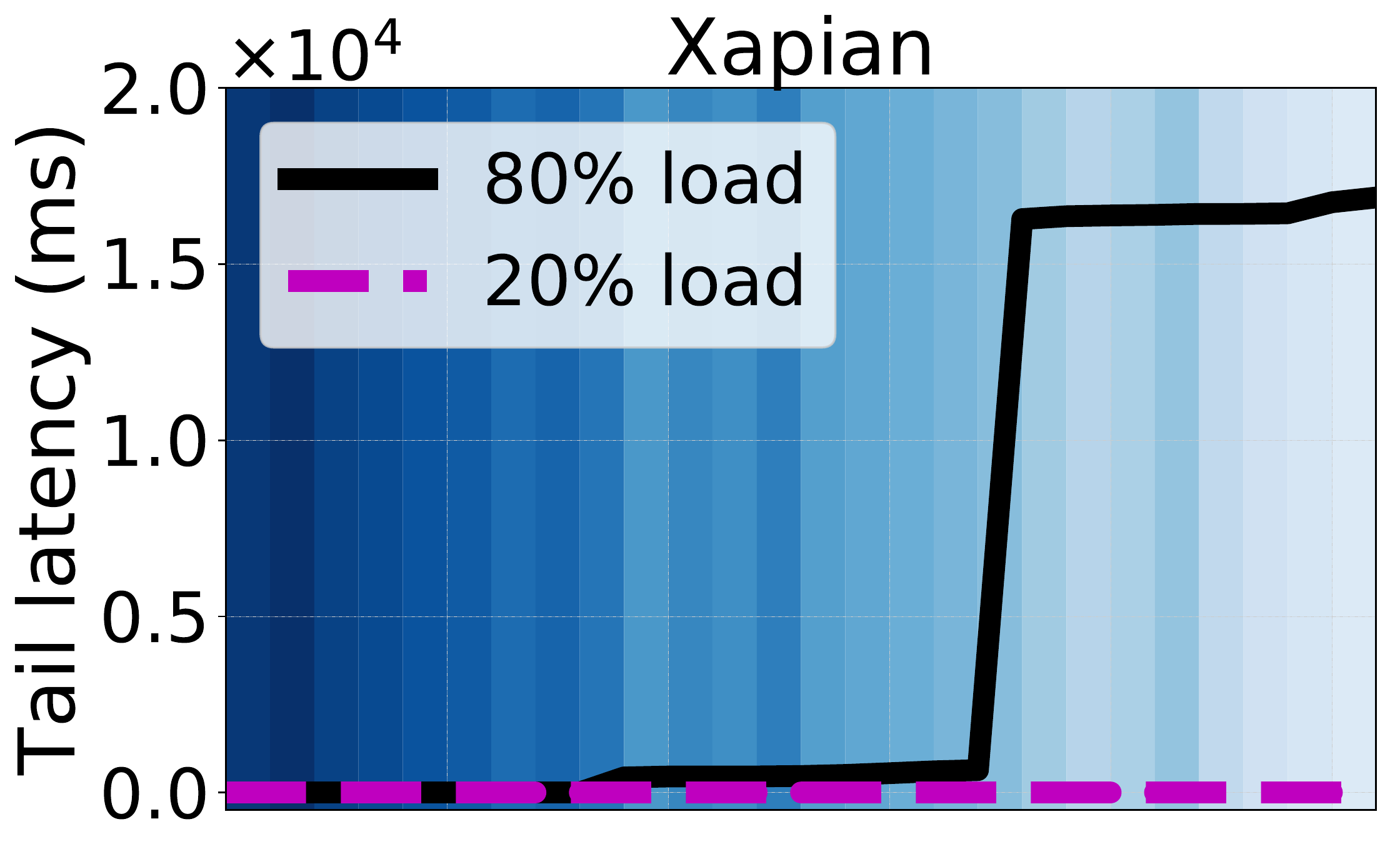} &
		\includegraphics[scale=0.166, bb=0 0 30 30,viewport= -110 0 480 390]{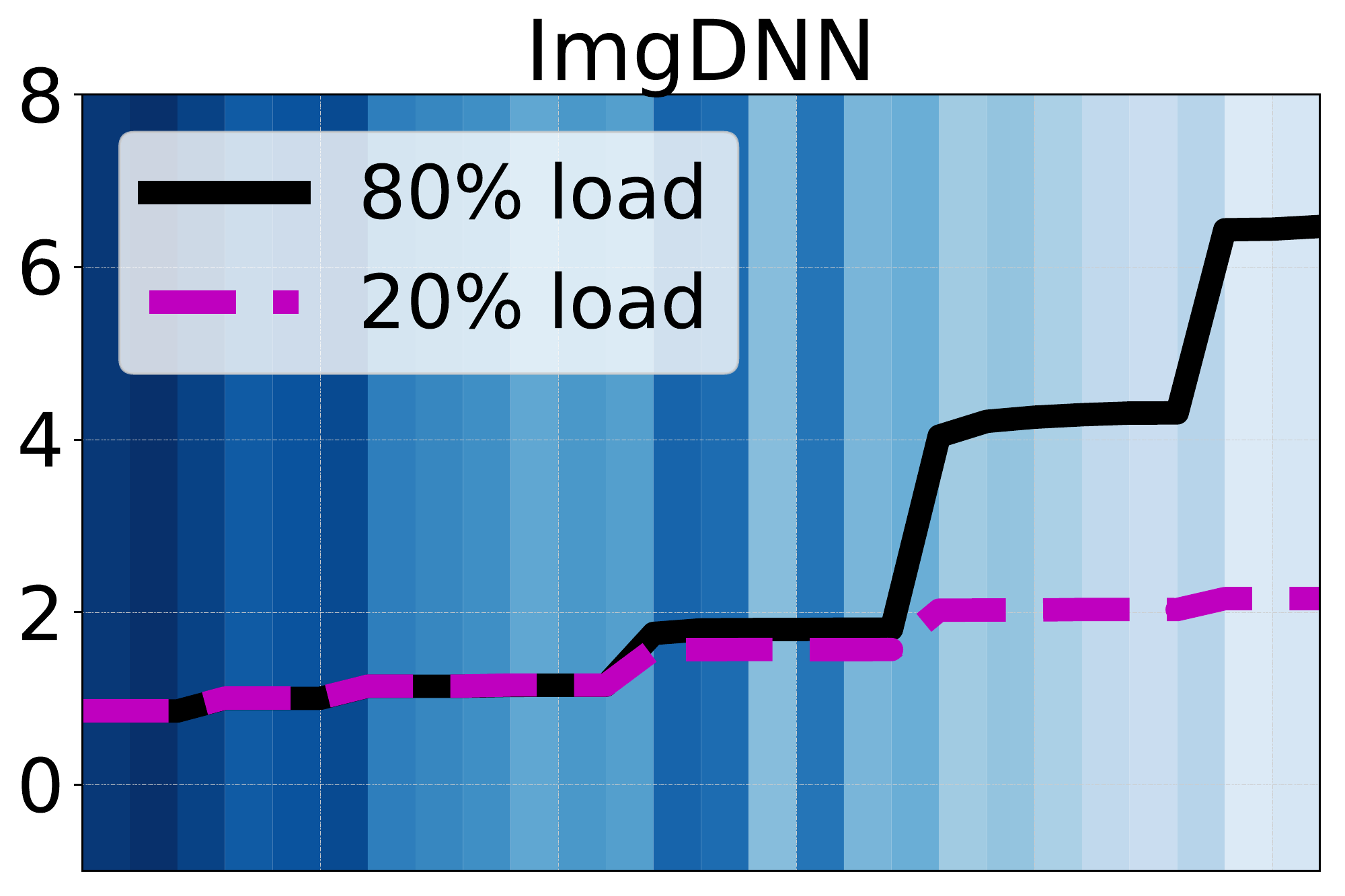} &
		\includegraphics[scale=0.166, bb=0 0 30 30,viewport= -60 0 480 390]{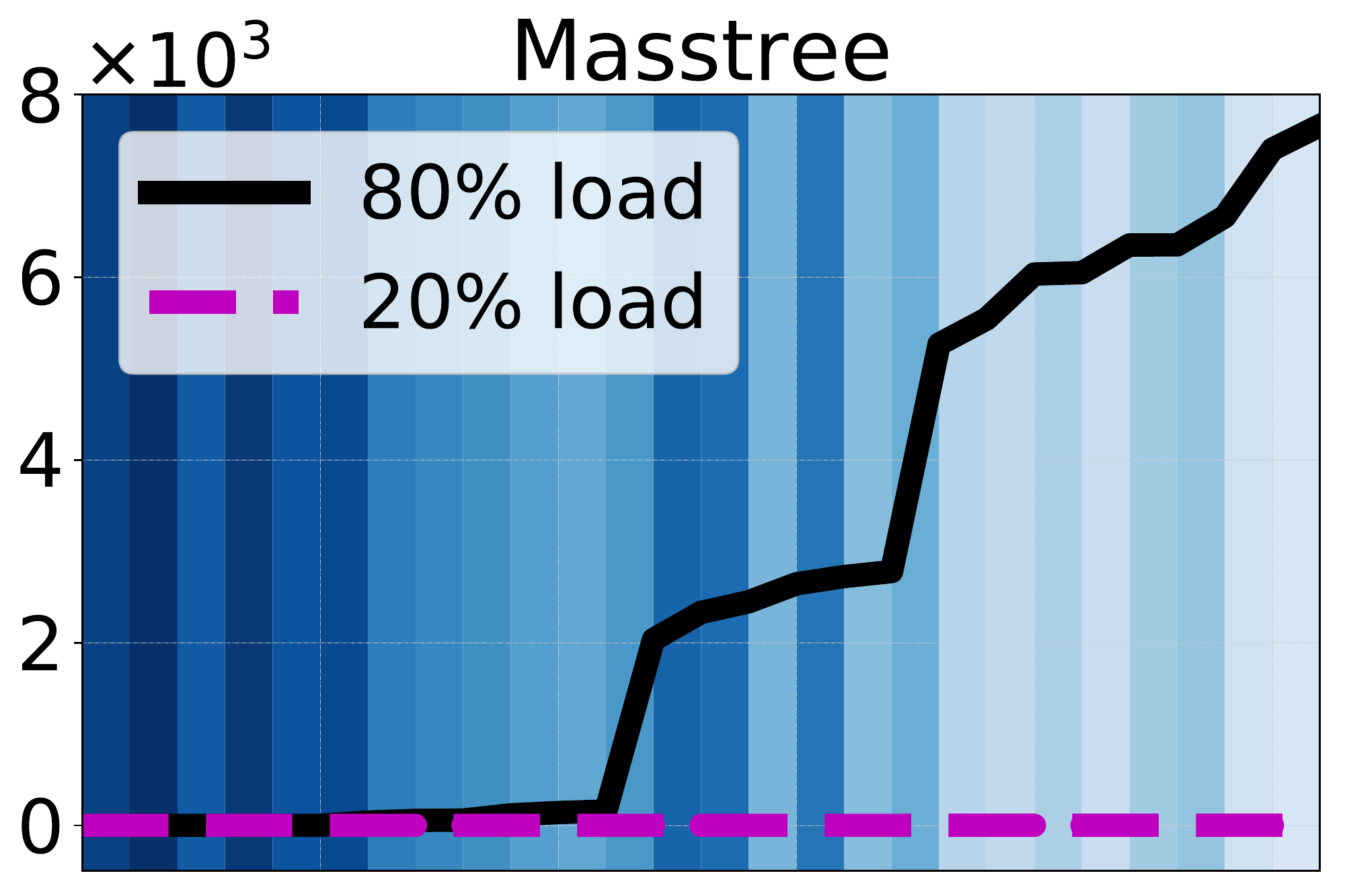} &
		\includegraphics[scale=0.166, bb=0 0 30 30,viewport= -50 0 480 390]{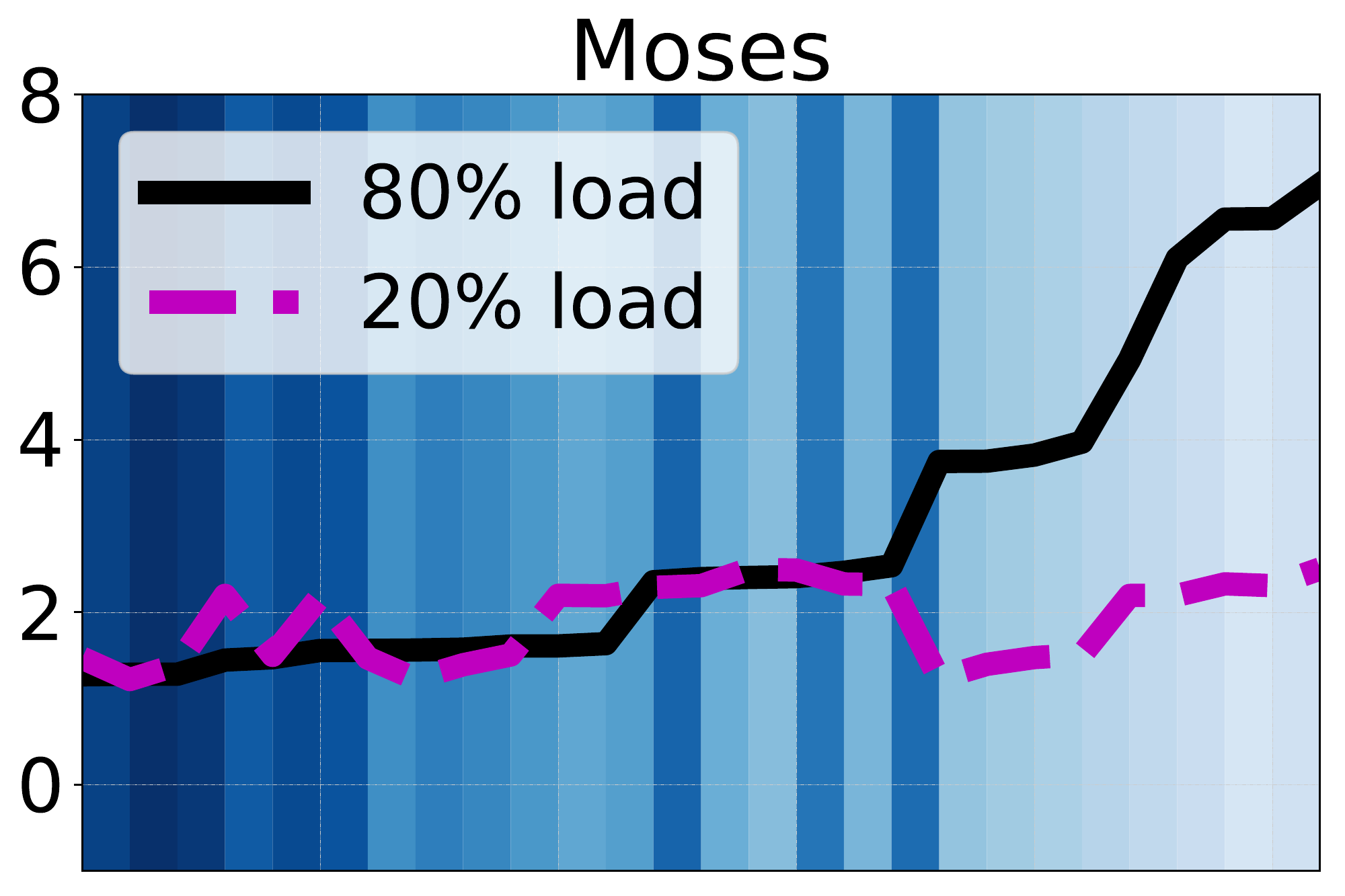} &
		\includegraphics[scale=0.166, bb=0 0 30 30,viewport= 550 0 480 390]{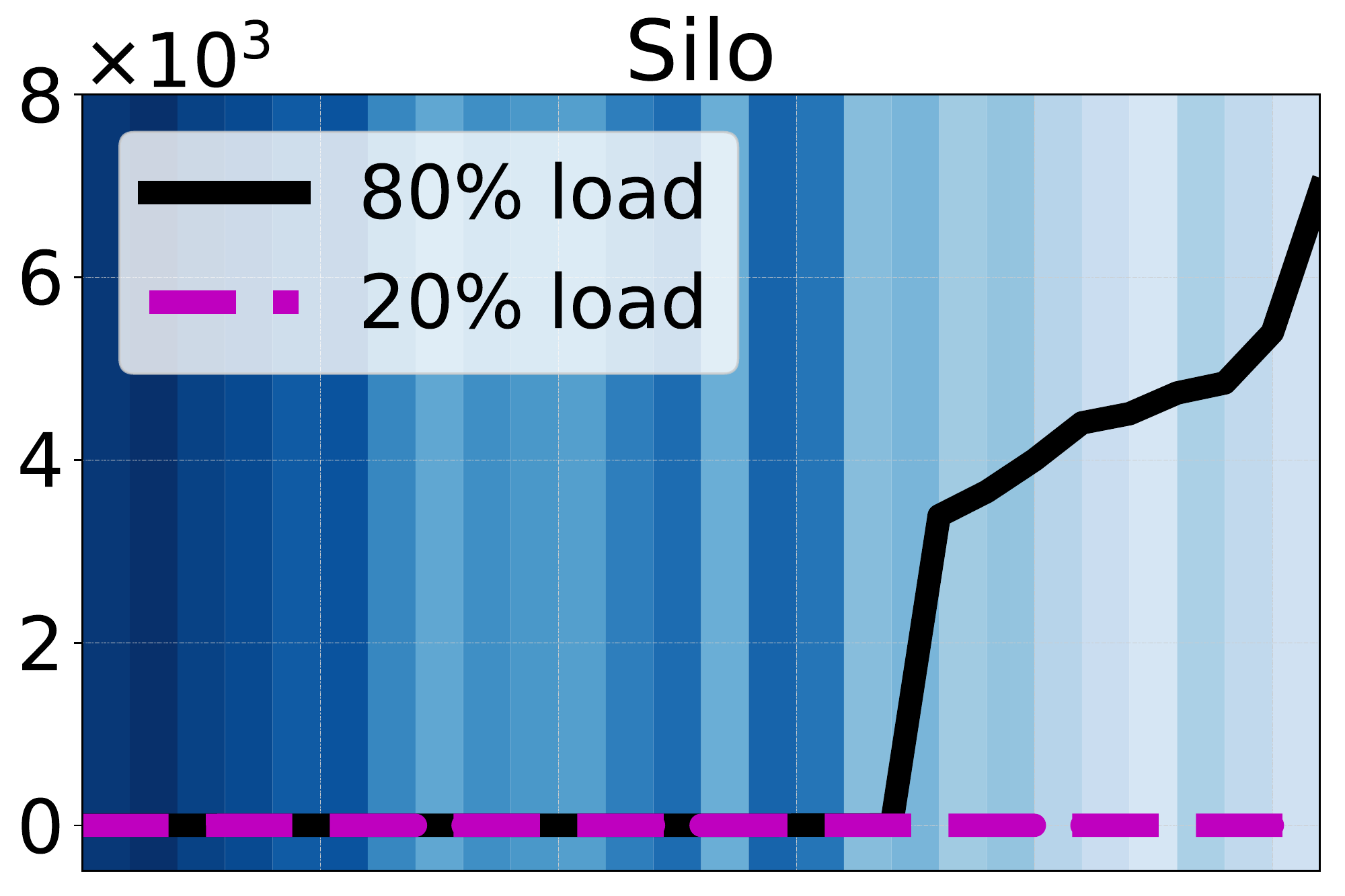} \\ 
		\includegraphics[scale=0.166, bb=0 0 30 30,viewport= 30 0 480 370]{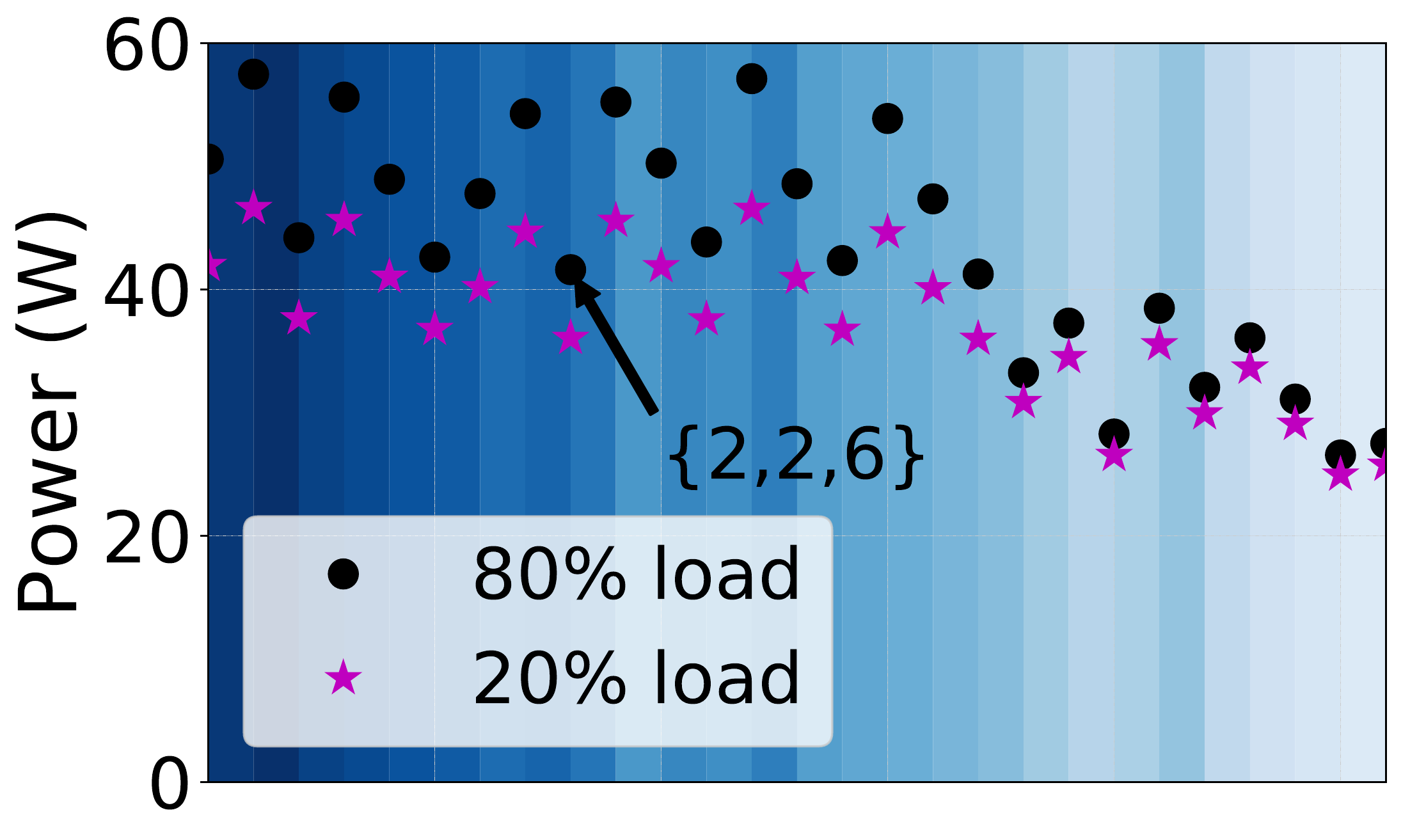} &
		\includegraphics[scale=0.166, bb=0 0 30 30,viewport= -70 0 480 370]{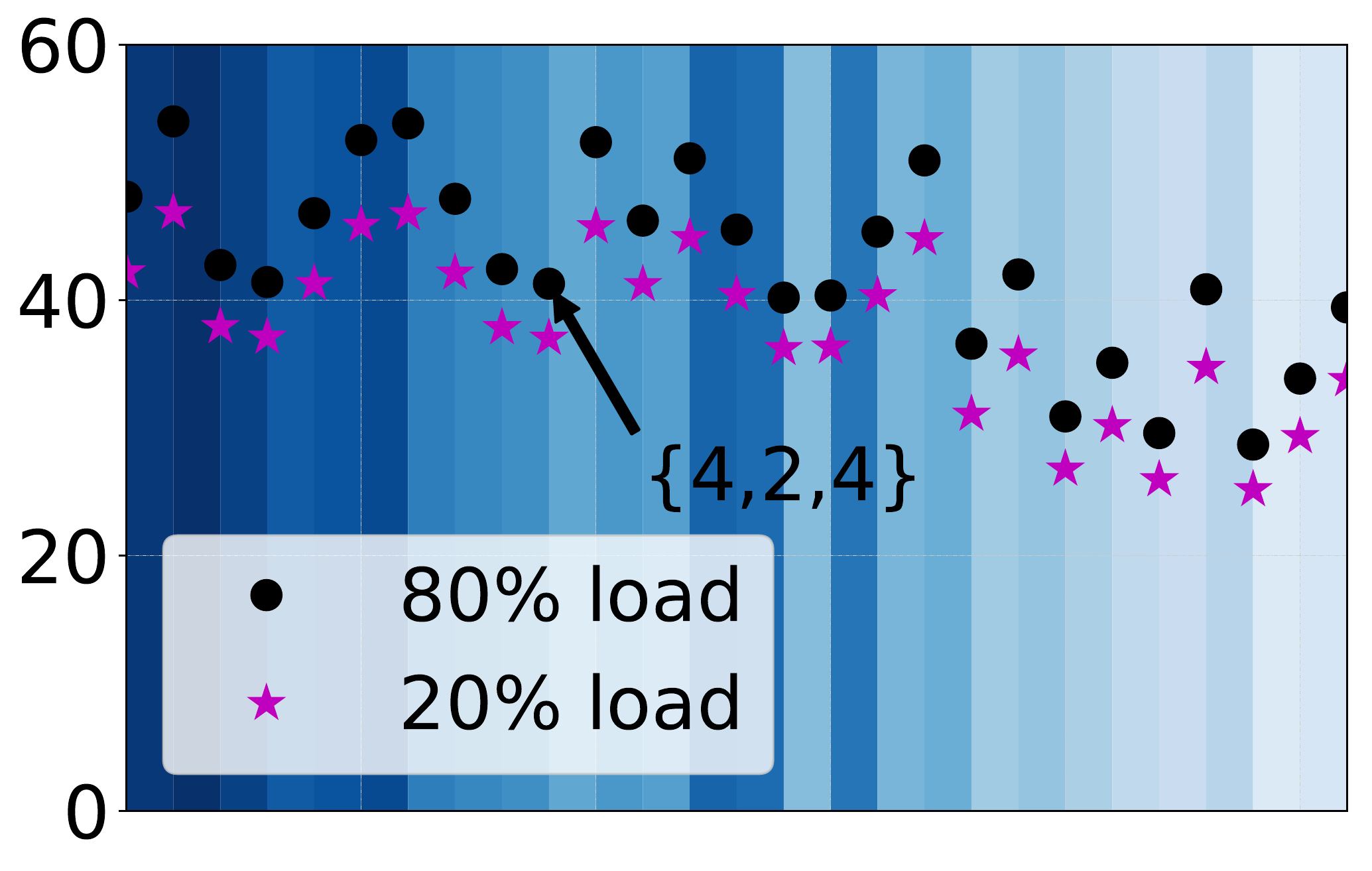} &
		\includegraphics[scale=0.166, bb=0 0 30 30,viewport= -20 0 480 370]{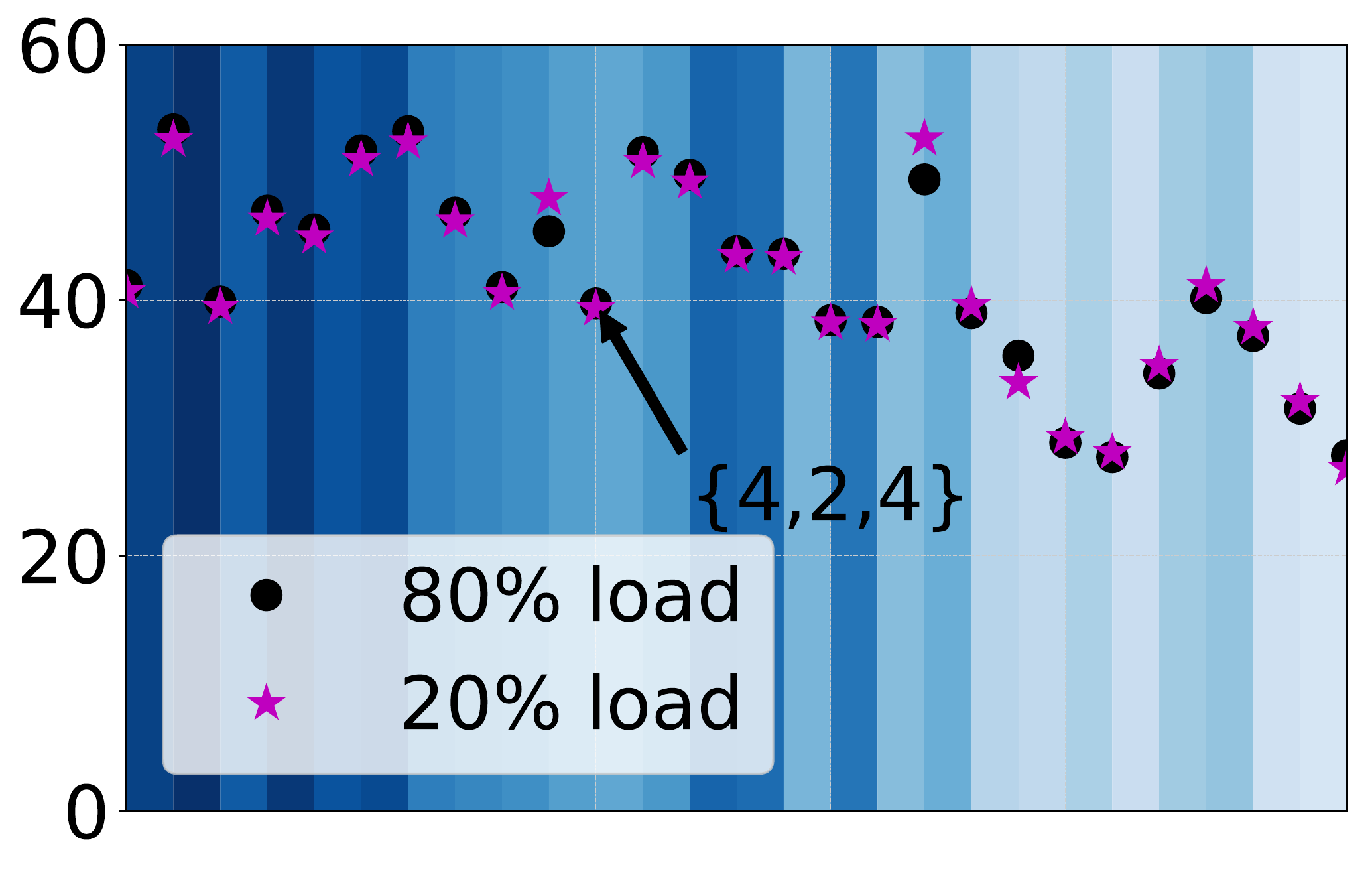} &
		\includegraphics[scale=0.166, bb=0 0 30 30,viewport= -10 0 480 370]{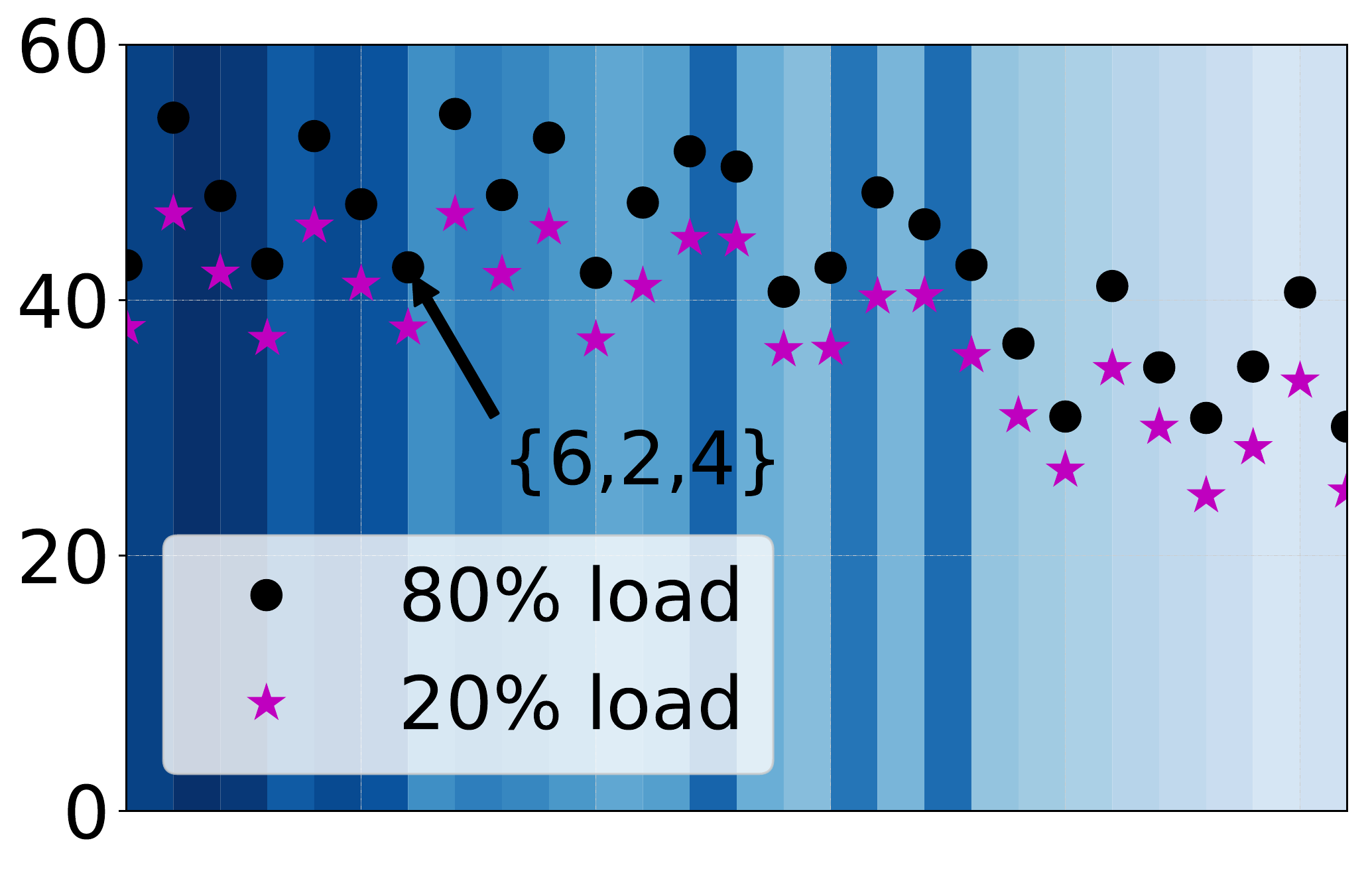} &
		\includegraphics[scale=0.166, bb=0 0 30 30,viewport= 590 0 480 370]{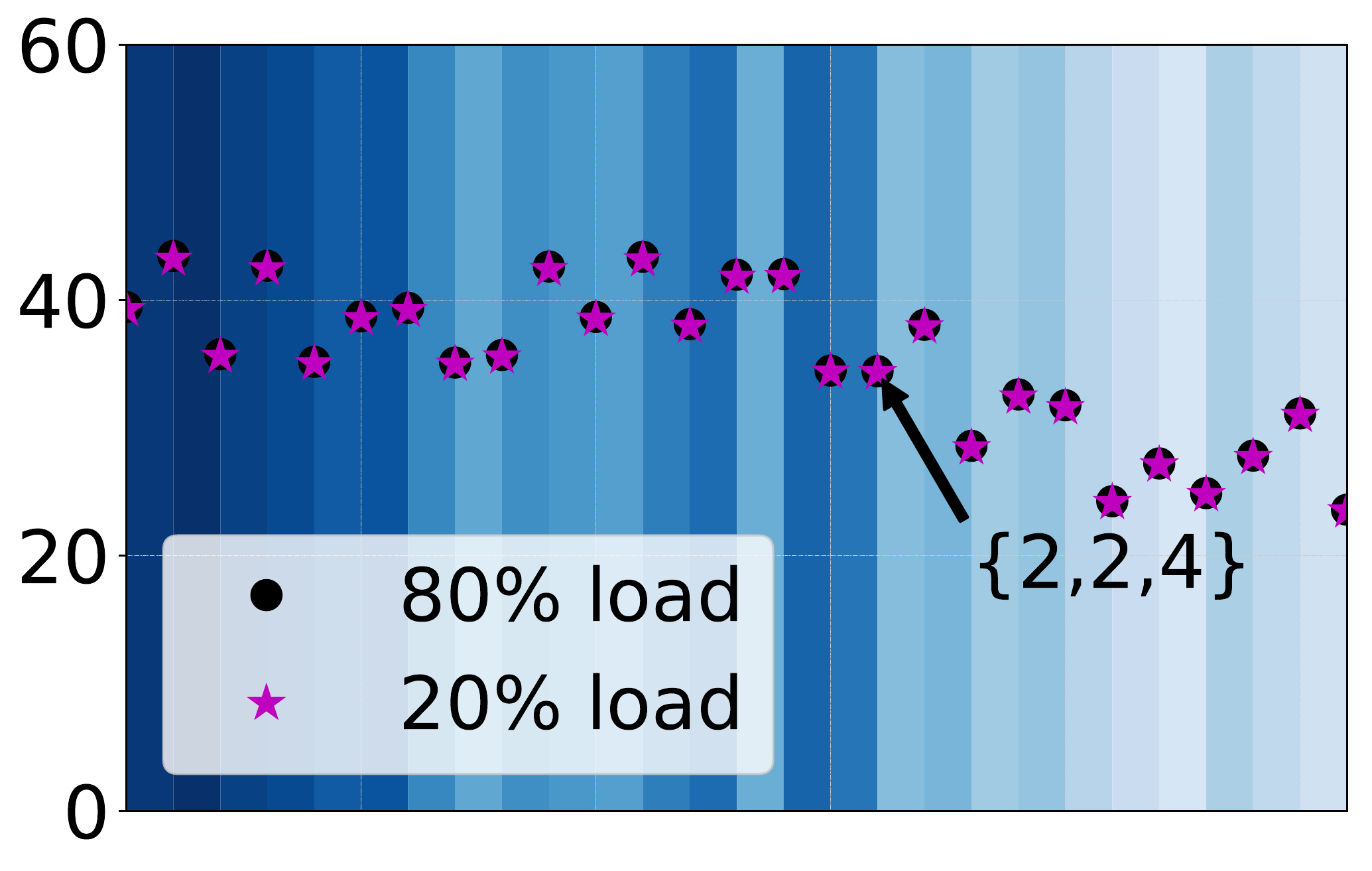} \\ 
		\multicolumn{5}{c}{\includegraphics[scale=0.25, bb=0 0 30 30,viewport= -450 20 1880 80]{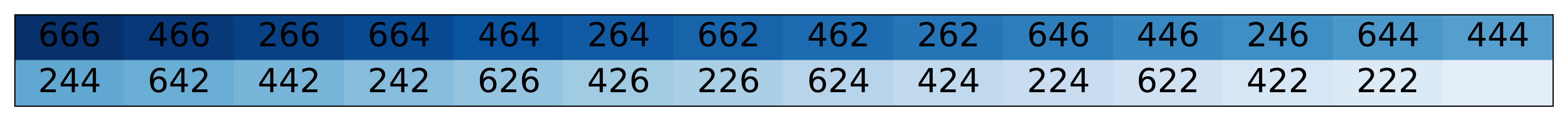}} \\ 
	\end{tabular}
  \caption{\label{fig:charac} {Characterization of tail latency and power of 5 latency-sensitive applications across core configurations. 
      Colors in the background represent the different core configurations, labeled as \{FE,BE,LS\}, as shown in the table. 
  Core configurations, from highest to lowest configuration (dark to light color), are ordered by serially decreasing configurations in LS, FE, and BE. 
  For each application, x-axis (core configurations) is sorted according to the tail latency observed at 80\% load.}}
  \vspace{-0.06in}
\end{figure*}

Teodorescu \emph{et al.}~\cite{Teodorescu:2008} and Winter \emph{et al.}~\cite{Winter:2010} propose thread scheduling
and power management for heterogeneous systems. Teodorescu~\cite{Teodorescu:2008} proposes LinOpt,
a linear programming-based approach, while~\cite{Winter:2010} explores the Hungarian algorithm to optimize
performance under a power budget. Adileh \emph{et al.}~\cite{AdilehHoles,DPDP} maximizes performance
by multiplexing applications between two voltage/frequency operating points to match the power budget.
The authors propose a technique to shift ``power holes'' arising due to core heterogeneity. 
{Navada \emph{et al.} \mbox{\cite{non-monotic}} propose the use of non-monotic cores, each optimized for different 
instruction-level behavior, and steer applications on appropriate core types using bottleneck signatures. }
 
\noindent{\textbf{Latency Sensitive Workloads}}: 
Petrucci \emph{et al.}~\cite{OctopusMan} show that simply using asymmetric multicores without redesigning system software results in QoS violations.
They propose a controller that maps jobs to the least power-hungry processing resources that can
satisfy QoS by incrementally assigning more slower or faster cores until QoS is met. 
Ren \emph{et al.}~\cite{SlowFast1, SlowFast2} propose a query-level slow-to-fast scheduler, where 
short queries run on slower cores and longer queries are promoted to faster cores to reduce their service latency.
The latter work~\cite{SlowFast2} also theoretically proves the energy efficiency advantages of asymmetric multicores over homogeneous systems. 
All of these efforts assume that cores of the desired speed are always available, which is not realistic.
Haque \emph{et al.}~\cite{eetl} take into account the fact that there is a limited number of cores of 
each type. They combine asymmetric multicores with DVFS and implement the slow-to-fast scheduler of~\cite{SlowFast1, SlowFast2}.
However, asymmetric multicores have a fixed number of core types (generally two), 
while reconfigurable cores provide a finer granularity of heterogeneity, enabling fine-grained performance/power tuning.
We compare CuttleSys against an oracle-like asymmetric multicore in Section~\ref{sec:results}.
 
\vspace{-0.03in}
\subsection{Reconfigurable Architectures}

Previous work on reconfigurable cores focuses on batch, throughput-bound workloads. 
Lee \emph{et al.} show the efficiency advantages and limits of adapting microarchitecture parameters to workloads. 
Lukefahr \emph{et al.}~\cite{composite} propose Composite cores, which pair big and little compute engines, 
and save energy by running applications on the small core as much as possible, 
while still meeting performance requirements. Padmanabha \emph{et al.}~\cite{Padmanabha:2013} propose 
trace-based phase prediction for migration of applications in Composite cores. 

Chrysso~\cite{chrysso} proposes an integrated power manager that uses analytical power 
and performance models and global utility-based power allocation. The configuration space of a 
core in our work is significantly larger compared to Chrysso~\cite{chrysso}, which makes the optimization problem more complex. 
Resource Constrained Scaling (RCS)~\cite{RCS} also aims to maximize performance 
in power-constrained multicores. In RCS, the resources of a processor 
and the number of operating cores are scaled simultaneously, which means that 
the system can operate in only a few different configurations. 

Khubaib \emph{et al.}~\cite{morphcore} propose a core architecture that dynamically morphs 
from a single-threaded out-of-order to a multi-threaded in-order core.  
FlexCore~\cite{FlexCore} can similarly morph into 4-way out-of-order, 2-way out-of-order, 
or 2-way in-order cores at runtime. 
{Tarsa \emph{et al.} \mbox{\cite{intel_cpuadapt}} propose a post-silicon clustered CPU architecture that combines 2 out-of-order execution clusters,
which can operate as an 8-wide execution engine or a low-power 4-wide engine. } 

{The Sharing Architecture \mbox{\cite{Sharing}} and Core Fusion \mbox{\cite{corefusion}} combine multiple simple out-of-order cores to form larger out-of-order cores. 
CASH \mbox{\cite{CASH}} also advances the Sharing Architecture with a runtime to find the best configuration for a single application which minimizes cost and meets QoS, 
using control theory and Q-learning. 
CuttleSys accounts for the interference between multiple co-scheduled applications that must all meet performance guarantees, 
and can be applied to the Sharing Architecture to quickly explore the design space of resource slices 
when multiple applications are hosted on a multi-tenant server, and arrive at suitable per-job resources. } 

Zhang \emph{et al.}~\cite{DCS} and Petrica \emph{et al.}~\cite{flicker} propose cores that can be 
reconfigured by scaling datapath components to save energy beyond DVFS. The dynamic scheme in Flicker~\cite{flicker} 
optimizes performance for a homogeneous multicore with reconfigurable 
cores under a power budget. Zhang \emph{et al.}~\cite{DCS} also show that reconfigurable cores 
significantly extend the performance-energy pareto frontier provided by DVFS. 

However, these systems are limited to batch applications, and do not consider the implications of tail latency on core reconfiguration. 
Moreover, Zhang \emph{et al.}~\cite{DCS} only consider a single core running one application.
In Section~\ref{sec:flicker}, we discuss why Flicker cannot be applied directly in this setting, 
and provide a quantitative comparison between Flicker and CuttleSys.

\section{Characterization of \\Latency-Critical Services}
\label{sec:characterization}

We now quantify the impact of different core configurations 
on the tail latency of interactive cloud services. We use five 
applications, \texttt{Xapian, Masstree, Imgdnn, Silo, Moses}, and configure them based on the analysis in~\cite{tailbench}. 
We simulate each application on a homogeneous 16-core system using zsim~\cite{Sanchez13}, a fast and cycle-level simulator, 
combined with McPAT v1.3~\cite{mcpat} for a 22nm technology for power statistics. 
A core is divided into three sections, front-end (FE - fetch, decode, ROB, rename, dispatch), 
back-end (BE - issue queues, register files, functional), and load-store (LS - LD/ST queues), each of which can be configured to six-way, 
four-way, and two-way, similar to Flicker~\cite{flicker}, except that we adopt a more aggressive superscalar design. 
These cores dynamically power gate associated array structures in each pipeline region when the configuration is downsized. 

Fig.~\ref{fig:charac} shows the variation of tail latency and power for each service, across core configurations 
at low and high load. 
Across all services, at high load, tail latency increases dramatically as the back-end and load-store queue 
are constrained. On the other hand, at low load, tail latency remains low, even for 
the lower-performing configurations. Therefore, when load is low, interactive services can leverage 
reconfiguration to reduce their power consumption, without a performance penalty. 

We also observe that the core section that most affects tail latency varies between applications. 
For Xapian, tail latency is primarily determined by the load-store queue size, with low latency requiring a six-way queue. 
In the cases of ImgDNN, Silo, and Masstree, tail latencies are low when FE and LS are configured to six- 
or four-way, while in the case of Moses, tail latency primarily depends on the front-end core section. 

At high load, the configuration with the best performance-power trade-off varies across services. 
{For example, \texttt{Xapian} consumes the least power in a \{2,2,6\} configuration while keeping tail latency low, 
while for \texttt{ImgDNN}, \texttt{Masstree}, \texttt{Moses}, and \texttt{Silo}, 
configurations \{4,2,4\}, \{4,2,4\}, \{6,2,4\} and \{2,2,4\} consume the least power respectively.
This shows that different core configurations are indeed needed by diverse applications. 
Also, batch applications differ in preferences from latency-critical applications.
This variability across loads and applications highlights the need for 
practical runtimes that identify the best core configurations of each application online. }



\section{CuttleSys Overview}
\label{sec:prob}
\begin{figure}
	\centering
	\includegraphics[scale=0.52, bb=0 0 30 30,viewport= 90 510 480 700]{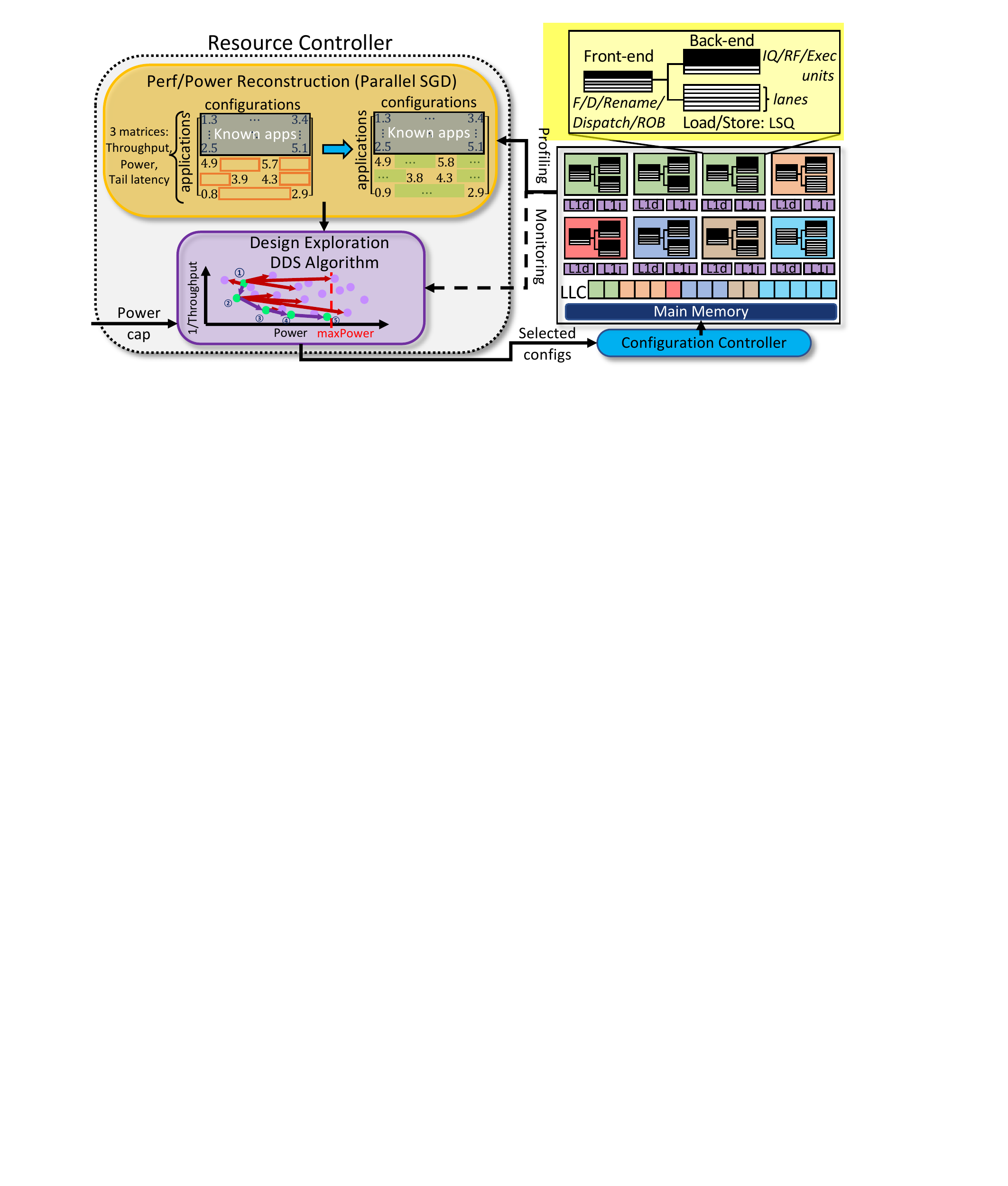}
  \caption{{CuttleSys system overview.} }
	\label{fig:block}
\end{figure}

We co-schedule latency-sensitive applications with batch workloads on a server with multiple reconfigurable cores, as shown in Figure~\ref{fig:block}. 
The last level cache (LLC) and power budget are shared across all cores. 

\vspace{-0.05in}
\subsection{Problem Formulation}

Our objective is to meet the QoS target for the latency-sensitive application, 
and maximize the throughput of the co-located batch applications, under 
a power budget that can change dynamically. 
Since the applications share the last level cache, the performance of each application depends on the interference in the last level cache caused by other applications. 
In order to mitigate this interference, CuttleSys also dynamically partitions the LLC among active applications at the granulariry of cache ways~\cite{Qureshi06,berkeley_isca}.



The system consists of $N$ cores. Each core can be configured in $m$ modes. Each application can be assigned one of $p$ cache way allocations.
Thus, each application can be executed in $m*p$ configurations.
For simplicity, the formulation below assumes one latency-sensitive application colocated with multiple ($B$) batch applications. 
The objective function is as follows:

\indent\indent$B_{i,{j,k}}$ = throughput (BIPS) of batch app $i$ running in core config $j$ and cache allocation $k$

\indent\indent{$T_{0,j,k}$ = tail latency of latency-sensitive app running in core config $j$ and cache allocation $k$}

\indent\indent{$P_{i,j}$ = power of app $i$ running in core config $j$}

\indent\indent{$C_{i,j,k}$ = cache ways allocated to app $i$ running in core config $j$ and cache allocation $k$}
\vspace{-0.08in}
\begin{align*}
\noindent I_{i,{j,k}} &= 1 \text{ if app $i$ is assigned to core configuration $j$} \\ &\text{       and cache allocation $k$} \\\vspace{-0.04in}
       \noindent &= 0 \text{ otherwise} \notag
\end{align*} 
\vspace{-0.16in}

We maximize the geometric mean of throughput:
\vspace{-0.08in}
\begin{align}\label{eq:GeoMean} 
  BIPS_{system} = (\prod_{i=1}^B\sum_{j,k} B_{i,{j,k}} * I_{i,{j,k}})^{1/B}
\end{align}
\vspace{-0.1in}
under the following constraints:
\vspace{-0.1in}

\begin{align}\label{eq:power} 
  {Power_{system} = \sum_{i=0}^B\sum_{j,k}  P_{i,j} * I_{i,j,k}} & {\leq maxPower}\\ 
\label{eq:cache}
\vspace{-0.1in}
{Cache\_alloc_{system} = \sum_{i=0}^B\sum_{j,k}C_{i,{j,k}}*I_{i,j,k}} & {\leq cacheWays}\\
\label{eq:qos}
\vspace{-0.1in}
{\sum_{j,k} T_{0,j,k} * I_{0,{j,k}}} & {\leq QoS}\\
\label{eq:assign}
\vspace{-0.1in}
\sum_{j,k} I_{i,{j,k}} = 1 & \: \forall \: i = 1, .. N
\end{align} 
Eq.~\ref{eq:power} states that the total power should be under the budget, while Eq.~\ref{eq:cache} states 
that the total allocated cache ways should be no higher than the LLC associativity. 
Eq.~\ref{eq:qos} addresses the QoS requirement of the latency-sensitive application. 
Eq.~\ref{eq:assign} states that each application can be mapped to a single configuration.
We use geometric mean as the objective function, since all batch applications have equal priority~\cite{geomean}.
Exhaustively exploring the full design space of core configurations and cache allocations ($(m*p)*(m*p)^{B}$) is impractical as the number of cores/applications increases. 
{This is problematic, since reconfiguration decisions need to happen online, and the optimization problem is non-linear and non-convex in nature.} 

Our scheme is made practical via two separate, mutually beneficial optimizations:
\begin{enumerate}
  \item Lightweight runtime characterization to infer  the performance, $B_{ i,{j,k}}$ in Eq.~\ref{eq:GeoMean}, $T_{0,j,k}$ in Eq.\ref{eq:qos} and power, $P_{i,j}$ in Eq.~\ref{eq:power}, of all applications across all possible $m$ core configurations and $p$ cache allocations; and
\item Fast and accurate design space exploration, given the output from (1) to determine a near-optimal solution  to the core configuration and cache allocation problem. 
\end{enumerate}


Previous approaches~\cite{flicker} to determine the impact of reconfiguration require detailed profiling of each active application 
against large number of resource configurations, 
which incurs non-trivial profiling overheads, and scales poorly with the number of configuration parameters. 
This approach is furthermore limited to batch applications, and does not take into account inter-application interference. 
Instead, we propose to infer performance (tail latency for interactive services and throughput for batch jobs) and power, 
across all possible core and cache configurations, by uncovering the similarities between the behavior of new and previously-seen applications across configurations. 
Specifically, we use PQ-reconstruction with Stochastic Gradient Descent~\cite{Witten,Delimitrou13, sgd1, sgd2}, a fast and accurate data mining technique that, given 
a few profiling samples for an application collected at runtime, estimates the application's performance and power across all remaining system configurations, 
based on how previously-seen, similar applications behaved on them. While SGD has been previously applied in the context of cluster scheduling~\cite{Delimitrou13,Delimitrou14}, 
core reconfiguration places much stricter timing constraints (only few ms) on SGD, as well as a larger configuration space, requiring a new, more efficient, parallel approximated SGD implementation. 


To quickly explore the design space, we adapt Dynamically Dimensioned Search (DDS)~\cite{DDS}, 
a heuristic algorithm that searches high-dimensional spaces for near-optimal solutions. DDS is computationally efficient, 
applicable to discrete problems, and especially effective for problems with high dimensionality, such as quickly searching 
the large space of resource configurations. The combination of  
SGD and DDS significantly improves performance over previous approaches.  

We also note that CuttleSys is an open-loop solution, which searches the design space and 
finds the best resource allocation in a single decision interval compared to feedback-based controllers, 
which take significant time to converge. This is especially beneficial for latency-critical applications, 
as they do not suffer from QoS violations until convergence. 

\subsection{Efficient Resource Management}
\label{sec:system-arch}


Fig.~\ref{fig:block} shows the high-level architecture of CuttleSys, 
which consists of the \emph{Configuration Controller} and the \emph{Resource Controller}. 
At the beginning of each decision quantum (100ms by default, consistent with prior work~\cite{flicker}), the \emph{Configuration Controller} 
profiles performance and power, which are used by the \emph{Perf/Power Reconstruction} module in the \emph{Resource Controller}. 
The \emph{Configuration Controller} then configures cores and cache ways based on the solution from the \emph{Design Exploration} module for the remainder of the timeslice.

The \emph{Resource Controller} takes as input the collected profiling samples, 
and the specified Power Cap, and determines the best core/cache configurations. 
The first step is \emph{Perf/Power Reconstruction}, 
which uses Stochastic Gradient Descent (SGD) to estimate 
the power and performance of an application for all core and cache configurations, 
based on a small number of samples (Section~\ref{sec:sgd}). 
The \emph{Design Exploration} uses SGD's output to determine the best
configuration for each job (Section~\ref{sec:dds}). 

\begin{figure}
	\centering
	\begin{tabular}{c}
		{\includegraphics[width=0.92\linewidth,trim=0cm 0 0cm 0, clip=true]{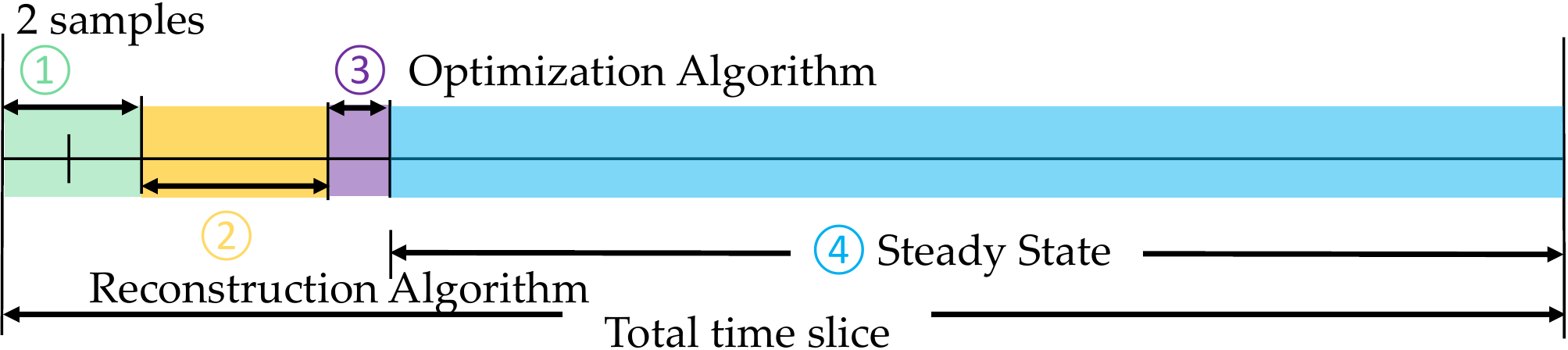}} \\
	\end{tabular}
	\caption{Timeline showing the steps of characterization, inference, and steady-state operation in CuttleSys. }
	\label{fig:timeline}
	\vspace{-0.12in}
\end{figure}

We describe the timeline of this process below, using Fig.~\ref{fig:timeline}.
Our approach requires $2$ profiling samples, one sample of the highest- and one 
of the lowest-performing configurations, corresponding to the widest-issue (\{6,6,6\}) and 
narrowest-issue (\{2,2,2\}) configurations respectively with one LLC way per core for the currently running applications, 
along with the performance and power of some ``training'' applications in all configurations, as shown in Figure \ref{fig:block}. 
We run applications for the duration 
of a sample timeframe ($1ms$ as described in Section~\ref{sec:profiling}), 
for each configuration and measure performance and power (\textcircled{1}). 
QoS for most cloud services is measured at intervals longer than $1ms$~\cite{Delimitrou14,agile12,Meisner11,Lo14,Lo15,Chen17}.
To obtain meaningful measurements, we measure tail latency over the entire $100ms$ of the previous timeslices. 
After this online profiling, we run the reconstruction algorithm to estimate the tail latency of latency-sensitive cloud services,  
the throughput of batch applications, and the power consumption of each application across all $m*p$ configurations (\textcircled{2}). 

Finally, we apply DDS to quickly search the space of core configurations and cache allocations, 
and find a solution that meets QoS and maximizes the throughput of batch applications for the given power budget (\textcircled{3}).
The system then runs in steady state (\textcircled{4}) with the selected core and LLC configurations. 
At the end of the timeslice, power and performance are measured and updated in the SGD matrix 
to ensure that any predictions deviating from the real metrics are corrected. 

\vspace{-0.08in}
\section{\hspace*{-0.05in}Practical Inference with SGD}
\label{sec:sgd}

The first step in the \emph{Resource Controller} estimates the power, throughput, 
and tail latency for applications across all core configurations and cache allocations. 
Previous techniques~\cite{flicker} require long profiling runs to accurately 
estimate an application's power and performance across configurations. 
Moreover, since previous work only targeted core configurations, estimating performance for cache allocations too would require an untenable number of profiling samples. 
Instead, we use the following insight to reduce profiling and improve practicality: the performance and power profile of a new, potentially unknown application 
may exhibit similarities with the characteristics of applications the system has previously seen, even if the exact applications are not the same. 

This problem is analogous to a recommender system~\cite{recommender,Netflix03,Bottou,Witten,Kiwiel,Gunawardana09,Burke02}, 
where the system recommends items to users based only on sparse information about their preferences. 
In our case, users are analogous to applications and items are analogous to resource configurations (combination of core configurations and cache allocations). 
A rating corresponds to the power or performance of an application running in the particular core and cache configuration. 
We construct a sparse matrix $R$ (one each for throughput, tail latency and power) with applications as rows and resource configurations (core-cache vectors) as columns. 
{The rows of matrix $R$ include some ``known'' applications, along the previously-unseen applications that arrive to the system. 
	The matrix is initially populated with the performance or power of these ``known'' applications 
which have been characterized once offline across all configurations.  
For all other new applications, the corresponding rows only have two entries obtained through profiling 
on two core-cache configurations out of the entire design space. 
The missing entries in the matrix are inferred using PQ-reconstruction with Stochastic Gradient Descent (SGD) 
\mbox{\cite{Delimitrou13, sgd1, sgd2, recommender, Netflix03}}.} 
{To reconstruct $R$, we first decompose it to matrices $P$ and $Q$, where the product of $Q$ and $P^{T}$ gives the reconstructed $R$, 
as shown in Algorithm}~\ref{alg:sgd}{. Matrices $Q$ and $P$ are then constructed using Singular Value Decomposition (SVD), and correspond to $Q = U$ 
and $P^{T} = \sum \cdot V^{T}$ respectively, where $U$, $V$ are the left and right matrices of singular vectors, and $\sum$ the diagonal matrix of singular values. 
In Algorithm}~\ref{alg:sgd}{, $A$ is the total number of applications (including known ones), and $m*p$ 
is the number of resource configurations.} {The impact of training set size is discussed in Sec.~{\ref{sec:evaluation_reconstruction}}}.
\vspace{-0.05in}

\begin{algorithm}
\caption{Reconstruction Algorithm}
\label{alg:sgd}
\begin{algorithmic}[1]
\State \textbf{Initialization}:
 \State \hspace{\algorithmicindent}$\mathbf{Q} \gets$ random(A,m*p); $\mathbf{P} \gets$ random(m*p,m*p) 
 \State \hspace{\algorithmicindent}$\eta \gets$ learning rate; $\lambda \gets$ regularization factor
 \State \hspace{\algorithmicindent} $maxIter \gets$ max \# of iterations
\For { $l \gets$ 1 to $maxIter$ }
    \For { $i \gets$ 1 to $A$ }
        \For { $j \gets$ 1 to $m*p$ }
            \State $\epsilon_{ij} \gets R_{ij} - \mathbf{Q_{j}} . \mathbf{P_{i}^{T}} $
            \State $\mathbf{Q_{j}} \gets \mathbf{Q_{j}} + \eta(\epsilon_{ij}\mathbf{P_{i}} - \lambda\mathbf{Q_{j}}) $
            \State $\mathbf{P_{i}} \gets \mathbf{P_{i}} + \eta(\epsilon_{ij}\mathbf{Q_{j}} - \lambda\mathbf{P_{i}}) $
        \EndFor
    \EndFor
\EndFor
\State $R \gets Q \times P^{T}$
\end{algorithmic}
\end{algorithm}

There is an obvious trade-off between the maximum number of iterations and the reconstruction accuracy: 
the fewer the iterations, the lower the overhead, 
but also the higher the prediction inaccuracy. We have conducted a sensitivity study 
to select convergence thresholds for SGD. To further reduce overheads, 
we have also limited the number of iterations. 

{For the currently-running applications, we obtain two samples of the highest- and lowest-performing core configurations with the ways equally allocated at runtime.
We also get additional samples for these applications by monitoring power, throughput, and tail latency for the configurations from previous steady states.} 
To predict the throughput and power for the remaining configurations ($m*p-2$, initially but fewer as we get more points from previous steady states)
and tail latency for the remaining configurations ($m*p-1$ initially), we run three instances 
of the reconstruction algorithm, one each for throughput, tail latency, and power. 
{We run these three reconstructions in parallel to minimize overheads.} 

To further accelerate reconstruction, we have implemented a parallel reconstruction algorithm 
that executes SGD without synchronization primitives~\cite{hogwild,hogwild1}. This introduces a small, upper-bounded inaccuracy 
(approximately 1\%), while improving its execution time by $3.5\times$. 

\section{\hspace*{-0.03in}Fast Design Exploration with DDS}
\label{sec:dds}


Once SGD recovers the missing performance and power of each job 
across all core configurations and cache allocations, the system employs Dynamically Dimensioned Search (DDS) to quickly 
explore the space, and select appropriate core configurations and cache partitions. 
DDS~\cite{DDS} is specifically design to navigate spaces with high dimensionality.

\begin{figure}
\centering
\includegraphics[width=0.92\linewidth,trim=0cm 0 0cm 0, clip=true]{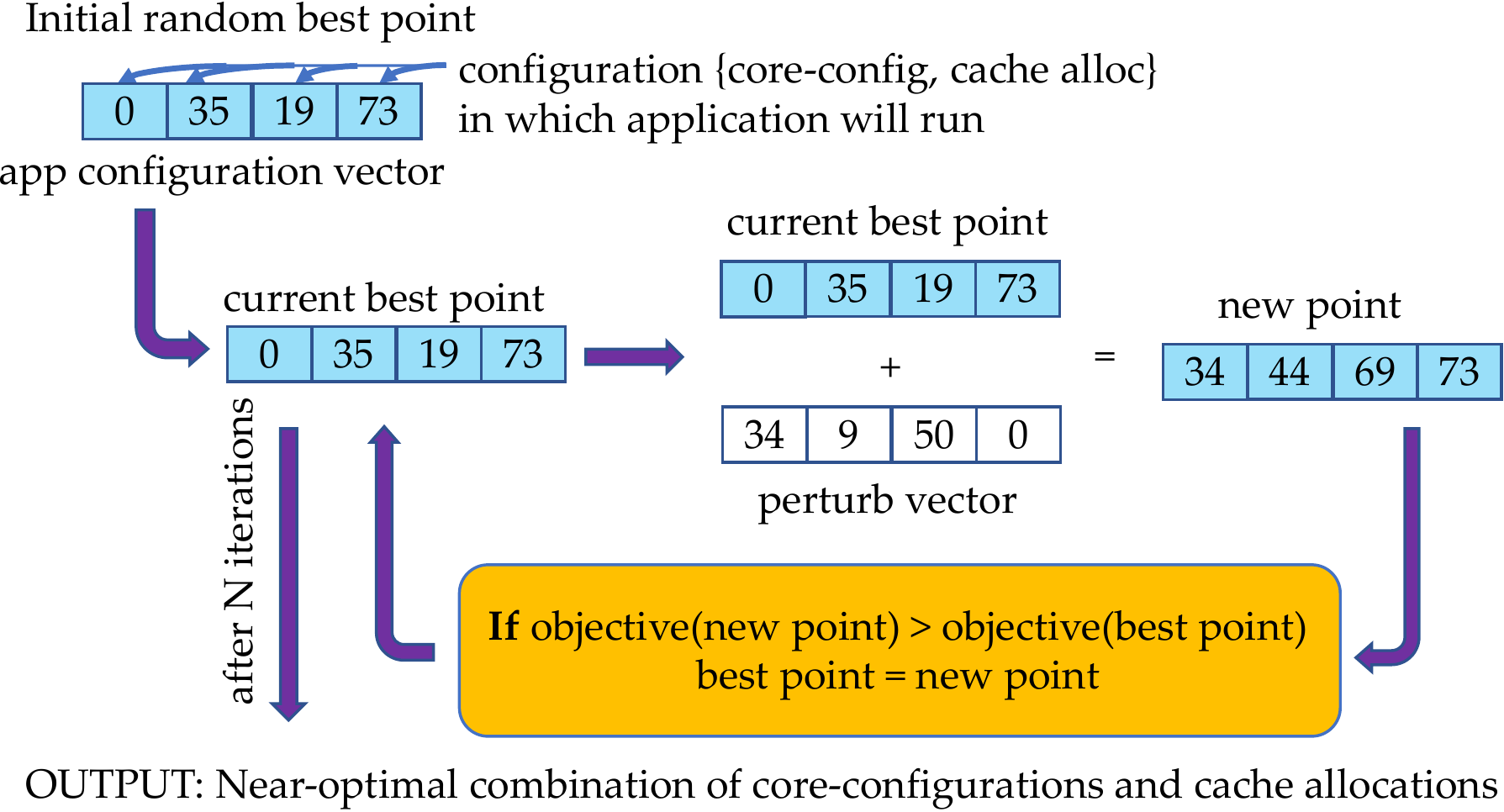}
\caption{The DDS design space exploration algorithm. }
\label{fig:dds-algo}
\end{figure}


%

The operation of DDS is shown in Fig.~\ref{fig:dds-algo}. The algorithm explores new points 
in the design space by perturbing a small number of dimensions from the current best point in each iteration, 
with the number of perturbed dimensions decreasing as the search progresses, and eventually converging to the best solution. 
Fig.~\ref{fig:dds-algo} shows an example of DDS for a simple 4-core system running four applications on four cores. 
The application configuration vector is a $N$-dimensioned decision variable, 
where the $i^{th}$ dimension denotes the configuration assigned to the $i^{th}$ application. The configuration assigned can be any number from 0 to $m*p-1$. The algorithm starts with 
a set of random points, and selects the point that has the highest value for the target objective as the current 
best point. In the given example, the current best point has threads $0$, $1$, $2$ and $3$ assigned to configurations $0$, $35$, $19$, and $73$ respectively. 
The current best point is then perturbed to explore new points. 
If the new point has a higher objective, it replaces the previous best point, 
and the process repeats until the algorithm arrives at a near-optimal combination of core configurations and cache allocations. 
The perturbation vector determines the number of dimensions to be perturbed and the perturbation magnitude 
for each dimension. DDS searches across more dimensions in the beginning, and narrows down to fewer dimensions later.
The perturbation quantity is equal to $r\cdot(\#\:confs)\cdot\mathcal{N}(0,1)$, where $r$ is a perturbation parameter.

\subsection{Handling Optimization Constraints}

The optimization problem described in Sec.~\ref{sec:prob} has three constraints: 
a) power (Eq.~\ref{eq:power}), b) cache (Eq.~\ref{eq:cache}), and c) QoS (Eq.~\ref{eq:qos}). 

Since latency-critical applications are load-balanced, all cores assigned to them
run in the same configuration. This simplifies the search for a suitable core configuration 
to just scanning through the predicted tail latency values of the $m*p$ configurations. 
We select the lowest cache allocation and the core configuration that consumes the least power while meeting QoS. 
DDS then explores points for the batch applications, while keeping the configuration 
of cores and cache ways assigned to latency-critical application fixed.

To handle the power and cache constraints of Eq.~\ref{eq:power} and~\ref{eq:cache}, 
we use an objective function that penalizes the points that consume more power and/or more cache than allowed as follows:
\begin{align*}
o&bjective(\mathbf{x}) =  BIPS_{system}(\mathbf{x}) \\
& - penalty\_power * ( maxPower - Power_{system}(\mathbf{x}) ) \\
& - penalty\_cache * ( maxWays - Cache\_alloc_{system}(\mathbf{x}) )
\end{align*}
We choose a soft penalty approach to handle the power constraint in the objective function, 
so that points with slightly higher power are not heavily penalized. 

If no configurations are found which meet the QoS of the latency-critical service, CuttleSys reclaims cores from 
the batch workloads, one per timeslice, and yields them to the latency-critical service, until QoS is met. 
The cores are similarly incrementally relinquished by the latency-critical applications when QoS is met with latency slack. 

\subsection{Parallel DDS}

To further speed up the design space exploration, we
have designed a new parallel DDS, shown in Alg.~\ref{alg:parallel-DDS}. 

\begin{algorithm}[h]
	\caption{Parallel DDS Algorithm}
	\label{alg:parallel-DDS}
	\begin{algorithmic}[1]
		\State \textbf{Initialization}:
		\State\hspace{\algorithmicindent} $maxIter \gets$ max \# of iterations
		\State\hspace{\algorithmicindent} $\textbf{r} \gets$ perturbation parameter 
		\State $l_c$ = get\_config\_LC()
		\State Initial rand points $\mathbf{x}$=$\{l_c, .., l_c, x_{K}, ..., x_N\}$
		\State $\mathbf{x^{best}} \gets arg max\{obj(\mathbf{x}) | \mathbf{x} \in random\:points\}$
		\For { $i \gets$1 to $maxIter$ }
		\State $\mathbf{x}^{localbest} = \mathbf{x}^{best}$
		\For { $j \gets$ 1 to $pointsPerIteration$ }
		\State $p \gets 1 - log (i)/log(maxIter) $
    \State add dimensions to $\{P\}$ with probability p
		\For {$ d \in \{P\}$ }
    \State $x^{new}\lbrack d\rbrack =x^{localbest} \lbrack d \rbrack +r\cdot(\# \: confs)\cdot\mathbf{\mathcal{N}}(0,1)$     
    \If {$x^{new} \lbrack d \rbrack \not\in [0, \# \: confs)$}
		\State reflect the perturbation
		\EndIf
    \EndFor
		\If { $obj(\mathbf{x}^{new}) > obj(\mathbf{x}^{localbest})$ }
		\State $\mathbf{x}^{localbest} = \mathbf{x}^{new}$
		\EndIf    
		\EndFor
		\State $barrier\_wait()$
		\If {$threadID == 0$}
		\State $\mathbf{x}^{best} \gets arg max\{obj(\mathbf{x}) | \mathbf{x} \in \{\mathbf{x}^{localbest}\}\}$
		
		\EndIf
		\State $barrier\_wait()$
		\EndFor
	\end{algorithmic}
\end{algorithm}

In the first phase, we initialize the algorithm's parameters. 
Line $2$ sets the maximum number of iterations ($maxIter$) of the algorithm. 
As $maxIter$ increases, the quality of the solution obtained improves, but at the same time 
the time required to run the algorithm also increases. 
We explore this trade-off in Section~\ref{sec:results}, and select the appropriate number of iterations. 

In parallel DDS, to avoid different threads exploring the same points (obtained from perturbation 
of the same best point), and to explore a larger space of configurations, we use four different values for the 
perturbation parameter; \textbf{r} = (r1, r2, r3, r4). In an $N$-core system, the first $N/4$ threads of the parallel algorithm set $r$ = $r1$, 
the next $N/4$ threads set $r$ = $r2$, and so on. 

Line $4$ gets the resource configuration that satisfies the QoS for {latency-critical (LC)} applications.
Lines $5$-$6$ show the randomly-chosen points the algorithm starts with, selecting the best among them as the initial best point. 
In parallel DDS, for a current best point, each thread generates $pointsPerIteration$ number of new points, and finds the best point 
among them, as shown in Lines $9$-$17$. The number of dimensions to be perturbed is determined by the probability function, 
as seen on Lines $10$-$11$, while Line $13$ shows the quantity by which the dimensions are perturbed. 
If the value of a dimension in the newly-generated point is out of bounds, the algorithm mirrors the value about
the maximum or minimum bound, to bring the point back within the valid range (Lines $14$-$15$).

DDS chooses the new point as the next best point if $obj(\mathbf{x}^{new}) > obj(\mathbf{x}^{best})$ (Lines $16$-$17$). 
After each core has computed $points$ $PerIteration$ points, a single core aggregates all the per-core best points, picks the best one, 
and distributes the selected configuration to be used for the next iteration (Lines $18$-$21$). 
DDS concludes after $maxIter$ iterations, and returns the best combination of core configurations and LLC allocations.

If the power cap is not met even after operating all cores running batch jobs 
in the lowest configuration, we turn off cores, in descending order of power, until the power budget is met.
\section{Experimental Methodology}
\label{sec:arch}

We evaluate our approach on 32-core multicore architectures consisting of reconfigurable cores. 
The core's architectural parameters are shown in Table~\ref{table:architecturalParams}, and 
are scaled according to the selected core configuration similar to~\cite{flicker}. 
Since we assume six-, four-, and two-way in each of the front-end, back-end, and load/store queue section, we have a total of $3^3 = 27$ ($m$=27) configurations.
Our reconfigurable cores are also similar to the large cores in AnyCore~\cite{anycore}, which evaluates the performance-energy overheads of reconfiguration. 

\vspace{-0.02in}
\begin{table}[tbh]
	\begin{center}
		\footnotesize
		\begin{tabular}{c|c}
			ine
			ine
			\multirow{4}{*}{\bf Front end} & BP: gshare + bimodal, 64 entry RAS, 4KB BTB \\
			& 144 entry ROB \\
			& 6-wide fetch/decode/rename/retire \\ 
			ine
			& out-of-order, 6-wide issue/execute \\
			& 192 integer registers, 144 FP registers \\
			{\bf Execution} & 48 entry IQueue, Load Queue, Store Queue \\
			{\bf core} 	& 6 Integer ALUs, 2 FP ALU \\
			& 1 Int/FP Mult Unit, 1 Int/FP Div Unit \\ 
			ine
			& L1 I-Cache: 32KB, 2-way, 2 cycles \\ 
			{\bf Memory} & L1 D-Cache: 64KB, 2-way, 2 cycles \\ 
			{\bf heirarchy} & L2 Cache: 64MB, shared, 32-way, 20 cycles \\ 
                      & 200 cycle DRAM access latency \\
			ine
      {\bf Technology} & 22 nm technology, 0.8V Vdd, 4GHz frequency\\
			ine
			ine
		\end{tabular}
    \vspace{-0.05in}
		\caption{Configuration of the 32-core simulated system. }
		\label{table:architecturalParams}
	\end{center}
\end{table}
\vspace{-0.12in}

{Based on the RTL analysis of frequency, energy, area overheads in}~\cite{anycore}{, 
we assume 1.67\% frequency and 18\% energy penalty per cycle for our reconfigurable cores compared to fixed ones. }
Reconfigurable cores also consume 19\% higher area. {In our experiments, we consider fixed-power scenarios, 
where the power budget is kept constant across the designs (core gating of symmetric and asymmetric multicore, and reconfigurable cores).} 
Under the power-capped scenarios, even if more cores can be packed in fixed-core designs (core gating-based and asymmetric multicores), 
they cannot be turned on due to power constraints. The performance benefits of CuttleSys are achieved at the cost of 19\% more area.

\subsection{Simulation Infrastructure and Workloads}
\label{sec:methodology}

We use zsim~\cite{Sanchez13} to obtain performance statistics combined with McPAT v1.3~\cite{mcpat} for 22nm technology to obtain power statistics. 
\setlength{\thickmuskip}{2mu}
We simulate 32-core systems, with 50\% cores assigned to a latency sensitive application and 50\% cores are assigned to batch jobs at time $t=0$. 
The core allocation changes at runtime as needed. 
Batch applications are multi-programmed mixes chosen from SPECCPU2006 
(\texttt{perlbench, bzip2, gcc, mcf, cactusADM, namd, soplex, hmmer}, \texttt{libquantum, lbm, bwaves}, \texttt{zeusmp}, 
\texttt{leslie3d, milc}, \texttt{h264ref}, \texttt{sjeng}, \texttt{GemsFDTD}, \texttt{omnetpp}, \texttt{xalancbmk}, 
\texttt{sphinx3}, \texttt{astar}, \texttt{gromacs}, \texttt{gamess}, \texttt{gobmk}, \texttt{povray}, \texttt{specrand}, 
\texttt{calculix}, \texttt{wrf}), while the {latency-critical (LC)} services 
are selected from TailBench~\cite{tailbench} (\texttt{Xapian, Masstree, ImgDNN}, \texttt{Moses, Silo}). 
To examine diverse resource behaviors, we co-schedule each of the TailBench applications with 10 multiprogrammed (16-app) mixes from SPECCPU2006, for a total of 50 mixes. 
We use one LC service for simplicity, however, CuttleSys is generalizable to any number of LC and batch services, as long as the system is not oversubscribed. 

The reconstruction algorithm requires the power and performance of a small number of representative applications to be collected offline, on all core configurations and cache allocations. 
We randomly selected 16 (discussed in Section~\ref{sec:evaluation_reconstruction}) of the above SPECCPU2006 applications for offline training 
at the beginning, excluding significant platform redesigns. 
Each of the multiprogrammed workloads is constructed by randomly selecting one of the remaining SPECCPU2006 benchmarks to run on each core, 
to ensure no overlap between the training and testing datasets. Each SPECCPU2006 benchmark runs with the reference input dataset. 

To find the maximum load each Tailbench service can sustain, we simulate it on a 16-core system and incrementally 
increase the queries per second (QPS), until we observe saturation. We use the QPS at the knee-point before saturation 
as the maximum load to avoid the instability of saturation~\cite{Chen19}. These max QPS are:
a) \texttt{Xapian}: 22{\smallcapital kQPS},  
b) \texttt{Masstree}: 17{\smallcapital kQPS},
c) \texttt{ImgDNN}: 8{\smallcapital kQPS},
d) \texttt{Moses}: 8{\smallcapital kQPS}, and 
e) \texttt{Silo}: 24{\smallcapital kQPS}.

The system's maximum power is the average per-core power 
across all jobs on reconfigurable cores scaled to 32 cores. We evaluate 
the system across power caps. 

\subsection{Baseline Core-Level Gating} 

We compare our design with core-level gating as it is widely employed in current systems for power gating.
To meet QoS the cores running latency-sensitive 
applications are always turned on. 
To determine which cores to turn off, core gating requires 
estimations of the power and performance of all applications. To do this, 
we profile the applications for one \emph{sample\_time}. 
We explore the following approaches for selecting the cores to turn off: 
a) descending order of power; b) ascending order of power; c) ascending order of BIPSperWatt; 
and d) ascending order of BIPS. From our experiments, we found that turning off cores based 
on descending order of power achieves the best performance for core-level gating. When turning off 
the last core required to meet the power budget, we search among the active cores and gate 
the one that meets the power budget with the smallest slack. 
{We also consider core-gating with LLC way-partitioning using \mbox{\cite{Qureshi06}}, since the technique is already available in real cloud servers\mbox{\cite{Lo15}}; 
	the choice of cache partitioning is orthogonal to the techniques in CuttleSys. }

Quantitatively comparing against core-level gating using the geometric mean of throughput is problematic, 
since when a core is gated, fewer applications run to completion. 
Thus, we compare the total number of instructions (useful work) executed over the same amount of time. 

\subsection{Asymmetric Multicores}
\label{sec:asymm}

Asymmetric multicores, which comprise cores with different performance and energy characteristics, have been proposed 
as an alternative to homogeneous multicores in order to improve energy efficiency \cite{hetero1,hetero2,hetero3,hetero4,hetero5,hetero6,hetero7}. 
Heterogeneity allows each application to receive resources that are suitable to its requirements and thus, 
improve the overall throughput, while still operating under a power budget. 
In asymmetric multicores, each type of core (typically a high-end and a low-power core type~\cite{armbiglittle}), and the number of cores of each type are statically designed. 
In contrast, reconfigurable multicores allow for finer granularity of configuration 
by providing higher number of different core types. Furthermore the number of cores in each configuration can be decided at runtime. 

We compare CuttleSys with a heterogeneous system with two types of cores: big cores, equivalent to the \{6,6,6\} configuration, and small cores, equivalent to the \{2,2,2\} configuration. While  
typically the number of cores are statically fixed, we compare against an oracle-like system, which selects the best number 
of big and small cores that meets the QoS of latency-critical applications, and 
maximizes the throughput of batch applications under a given power budget. 
For the oracle system, we also ignore any scheduling overheads that the threads 
incur to migrate between cores of different types. 

\section{Evaluation}
\label{sec:results}

\subsection{CuttleSys Scheduling Overheads}
\label{sect:overheads}

CuttleSys incurs three types of overheads: (i) for the initial \textit{application profiling} 
that gives the controller a sparse signal of the application's characteristics, 
(ii) for the \textit{reconstruction algorithm} that infers performance and power 
on all non-profiled configurations, and (iii) for the DDS space exploration (Fig.~\ref{fig:timeline}). 
Table~\ref{table:overheads} shows these overheads. 

\begin{table} [h]
	\centering
	\begin{tabular}{cc|c|c}
		\multicolumn{2}{c|}{Performance/Power} & SGD & DDS \\
		\multicolumn{2}{c|}{sampling} & reconstruction & search \\
		\hline
		\hline
		\multicolumn{1}{c|}{Single run} & Total time & \multirow{2}{*}{4.8 ms} & \multirow{2}{*}{1.3 ms}\\
		\cline{1-2}
		\multicolumn{1}{c|}{1 ms} & 2 ms & & \\
	\end{tabular}
  \vspace{-0.08in}
\caption{Characterization and optimization overheads. }
\label{table:overheads}
\end{table}
\vspace{-0.2in}
\subsubsection{Profiling}
\label{sec:profiling}

We empirically set a monitoring period of $1ms$ as a advantageous trade-off between 
reducing profiling overheads and increasing decision accuracy, similar to~\cite{flicker}. 
We profile all cores in parallel for $2ms$ (\textcircled{1} of Fig.~\ref{fig:timeline}), 
$1ms$ each in the widest-issue \(\{6,6,6\}\) and narrowest-issue \(\{2,2,2\}\) configurations 
with one way of LLC allocated to each core, and measure performance and power consumption. 
To avoid power overshoot by running all cores in the highest configuration, 
half of the cores run in the widest-issue configuration, and the other half in the narrowest-issue configuration 
in the first $1ms$ and vice-versa in the second $1ms$. Note that even core-level gating 
incurs an overhead of $1ms$ for one profiling period.  

\subsubsection{Reconstruction Algorithm} 
\label{sec:evaluation_reconstruction}

{Reconstruction requires characterizing offline a few ``known'' applications. } 
We select the fewest jobs (16) needed to keep accuracy over 90\% for all running
applications. If instead the training set included 24 applications, 
the inaccuracy drops to 8\%, {while execution time of the reconstruction algorithm} increases by 18\%. On the other hand, decreasing 
the training set to 8 applications increases inaccuracy to 20\%. 

We run three instances of the reconstruction algorithm (one each for throughput of batch jobs, 
tail latency of latency-sensitive applications, and power for all jobs).
Reconstructing the throughput for batch jobs takes longer, as it needs to find
the missing values for all combinations of core and LLC configurations for 16 applications, while reconstructing the 
tail latency needs to estimate the missing values for all configurations of 1 application at a time. 
Inferring performance and power for all possible LLC allocations (32 in our case) 
increases the overhead and impacts accuracy, even though many allocations would not 
be feasible in practice, as all 32 cores need to share the 32 ways. 
Therefore, we limit the LLC allocations for each job to 1/2, 1, 2, and 4 ways. 
If two jobs are allocated 1/2 ways each, both are assigned the same LLC way.
Any interference between them is handled by updating the entries in the reconstruction matrix 
with the measured values during runtime. 
The three reconstructions all run in parallel on the same server. 

\begin{figure}
	\begin{tabular}{ccc}
    \subfloat[]{\includegraphics[scale=0.26, viewport= 60 20 200 260]{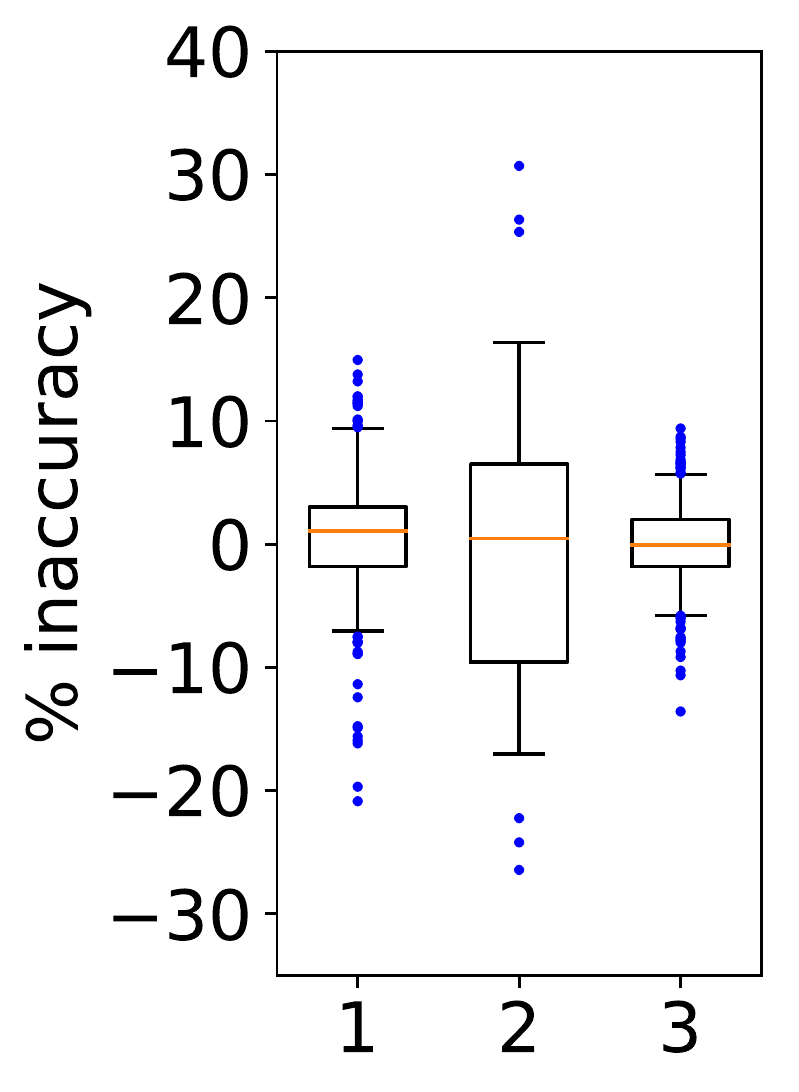}}
    &\subfloat[]{\includegraphics[scale=0.26, viewport= 60 20 200 260]{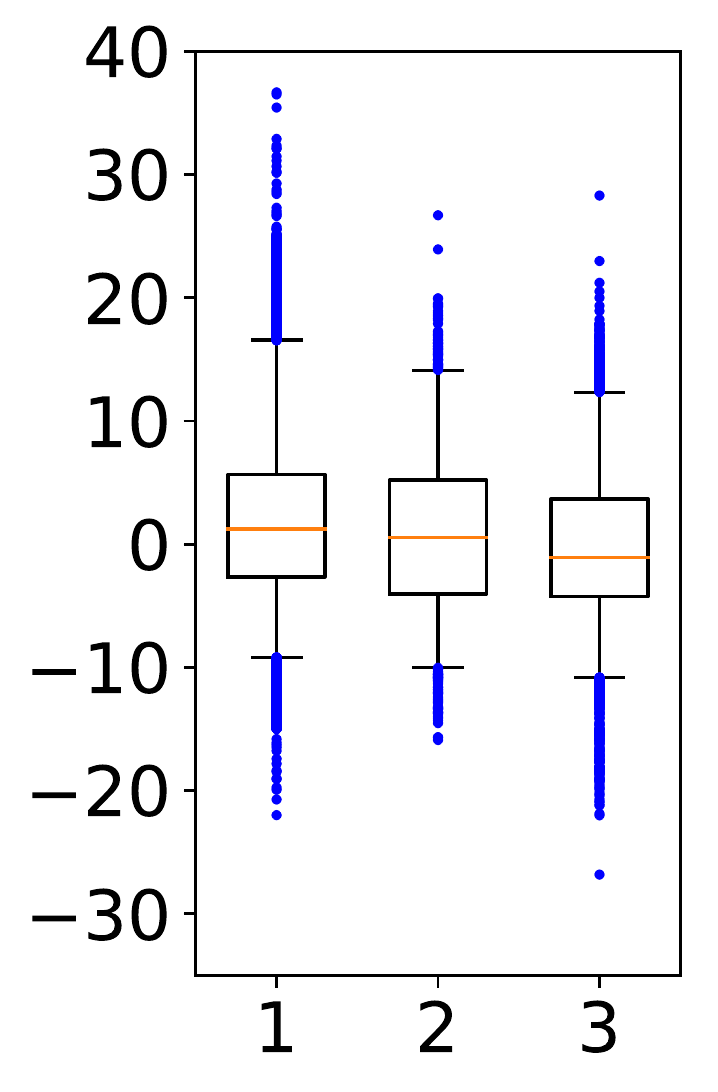}}
    &\multirow{2}{*}{\subfloat[]{\includegraphics[scale=0.28, viewport= 80 10 480 55]{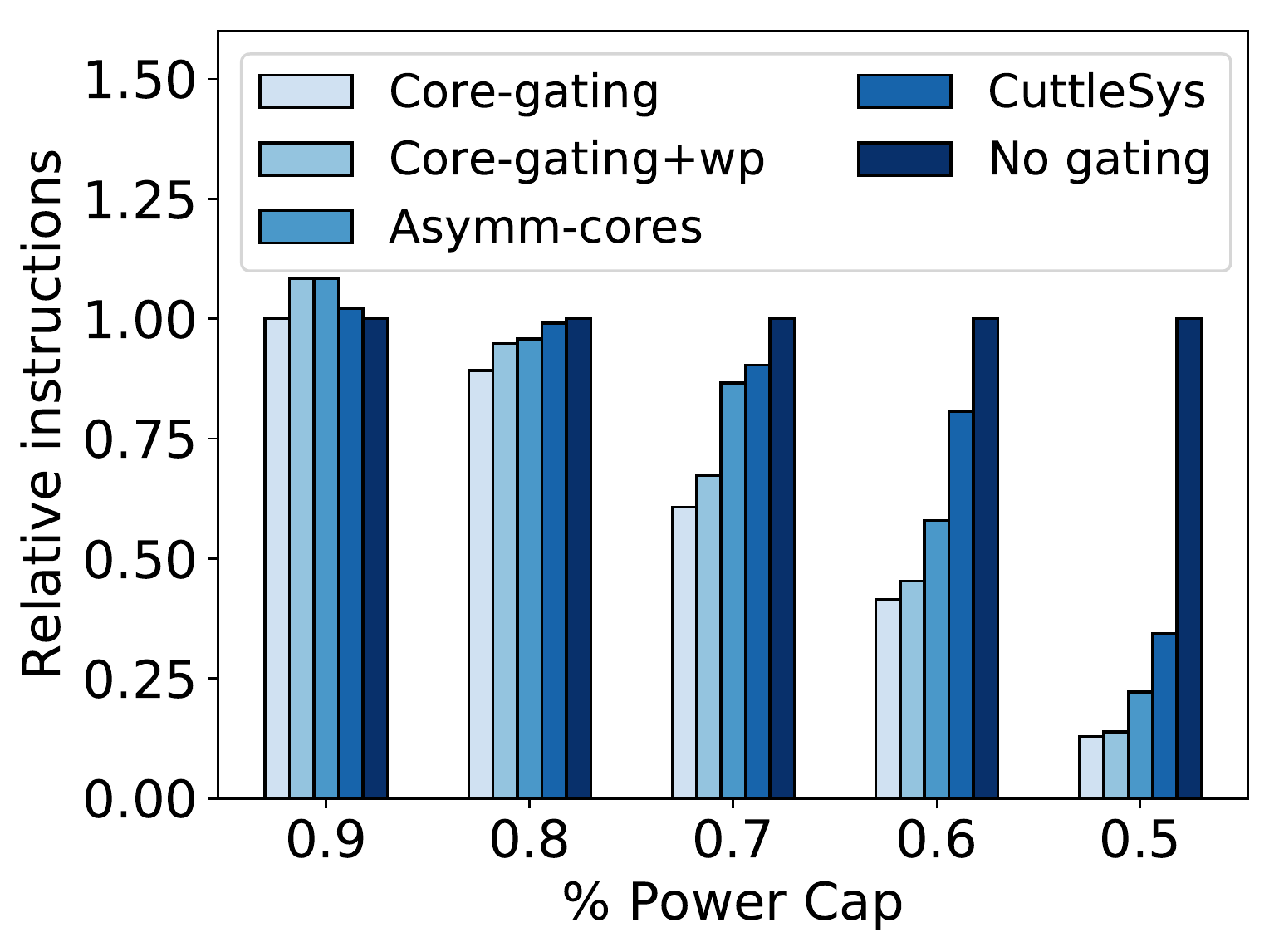}}}\\
  \multicolumn{2}{c}{\includegraphics[scale=0.287]{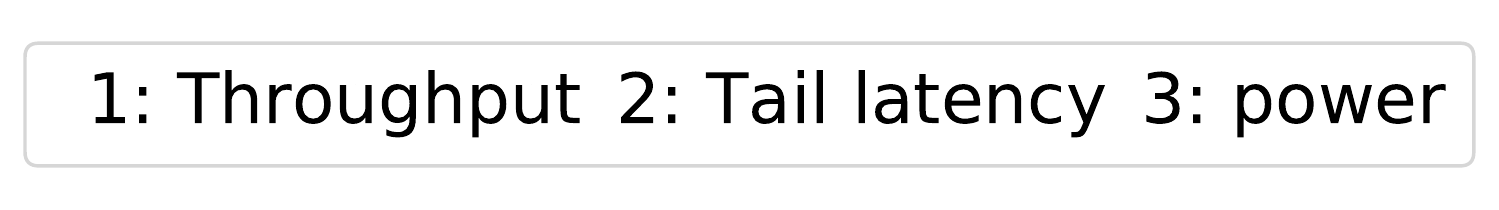}} & \\
  \end{tabular}
  \vspace{-0.16in}
	\caption{Box plots of the error between the measured and predicted performance and power by SGD 
		across configurations (a) in isolation and (b) with colocation. (c) Instructions with CuttleSys vs. 
core-level gating over $1s$ across power caps. }
	\label{fig:result}
  \vspace{-0.08in}
\end{figure}

\subsubsection{DDS Algorithm}

As described in Section~\ref{sec:dds}, the $\#\:confs$ is set to 107, 
since we consider four LLC allocations for each core configuration. 
We have performed sensitivity studies to find the parameters of parallel DDS that
achieve the best trade-off between runtime and accuracy.
We arrived at the parameter values shown in Figure~\ref{table:dds-params}.

\begin{wrapfigure}[7]{l}{0.27\textwidth}
	\vspace{-0.08in}
\footnotesize
\begin{tabular}{l|l}
ine
ine
initial random points & 50            \\ ine
\textbf{r}  = [r1,r2,r3,r4]  & [0.2,0.3,0.4,0.5] \\ ine
penalty\_wt      & 2                  \\ ine
pointPerIteration & 10 \\ ine
maxIter & 40 \\ 
ine
ine
\end{tabular}
  \vspace{-0.12in}
\caption{DDS parameters. }
\label{table:dds-params}
\end{wrapfigure}

\subsection{CuttleSys Inference Accuracy}
\label{sect:runtime-accuracy}

CuttleSys uses three instances of the parallel SGD algorithm
to reconstruct the throughput, tail latency, and power of co-scheduled applications 
across resource configurations. 

To isolate the prediction accuracy of SGD, we run all test applications in isolation 
for the full time slice in all core configurations, which avoids both interference from co-scheduled jobs and 
inaccuracies from limited profiling time. 
For the throughput, power, and tail latency estimation, we profile on two configurations per job, 
and infer the remaining 106 entries. 
Fig.~\ref{fig:result}(a) shows the estimation errors for throughput, tail latency, 
and power across the 12 ``testing'' SPEC applications 
and 5 Tailbench applications at 80\% load. Fig.~\ref{fig:charac} shows that 
some configurations incur very high tail latency, and are not selected during runtime. For these configurations, 
exact latency prediction is less critical, as long as the prediction shows that QoS is violated. 
We observe that the $25^{th}$ and $75^{th}$ percentiles are within 10\%, 
while the $5^{th}$ and $95^{th}$ percentiles are less than 20\% for throughput, tail latency, and power.
The error for tail latency is higher, as we predict services one at a time and only use 2 sample runs 
to predict the remaining 106 configurations.

We now examine the inaccuracy at runtime, which also includes the inter-application interference and the inaccuracies due to limited profiling time. 
Fig.~\ref{fig:result}(b) shows box plots of these errors for throughput, tail latency, and power. 
The median is close to zero and the $25^{th}$ and $75^{th}$ percentiles are within 10\% in all cases. 
However, the $5^{th}$ and $95^{th}$ percentiles in the case of tail latency increase, as do the outliers for throughput. 
This is due to (a) applications changing execution phases, making the profiling runs not representative of steady state behavior, 
and (b) resource contention between co-scheduled applications. 
Since CuttleSys updates the reconstruction matrix with the measured metrics, it accounts for changes at runtime. 

\begin{figure}
\begin{tabular}{ccc}
	\hspace{-0.4in}\subfloat[Core gating]{\includegraphics[scale=0.24, viewport= -40 0 480 350]{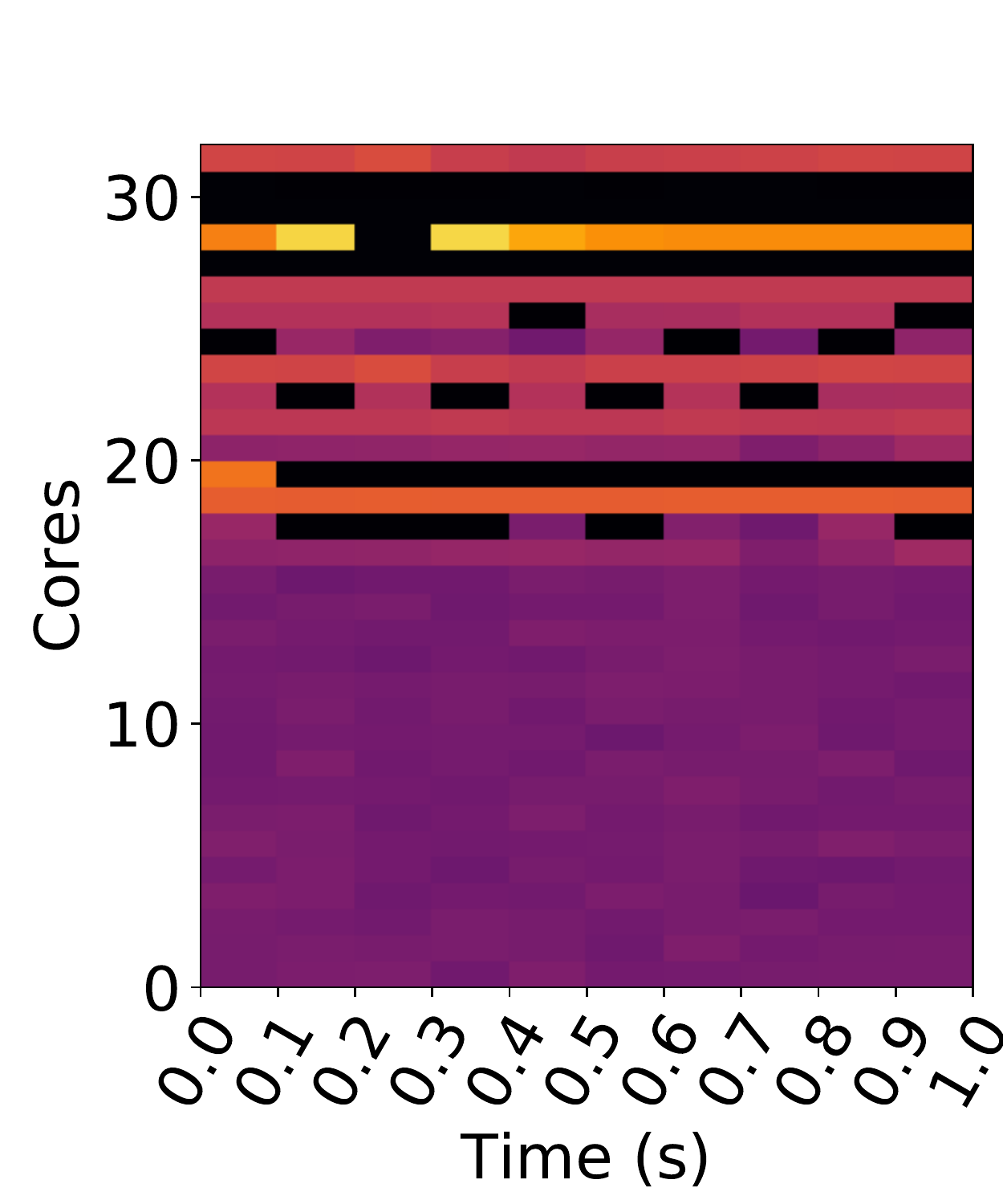}}
	&\hspace{-0.98in}\subfloat[Asymm Cores]{\includegraphics[scale=0.24, viewport= -150 0 480 350]{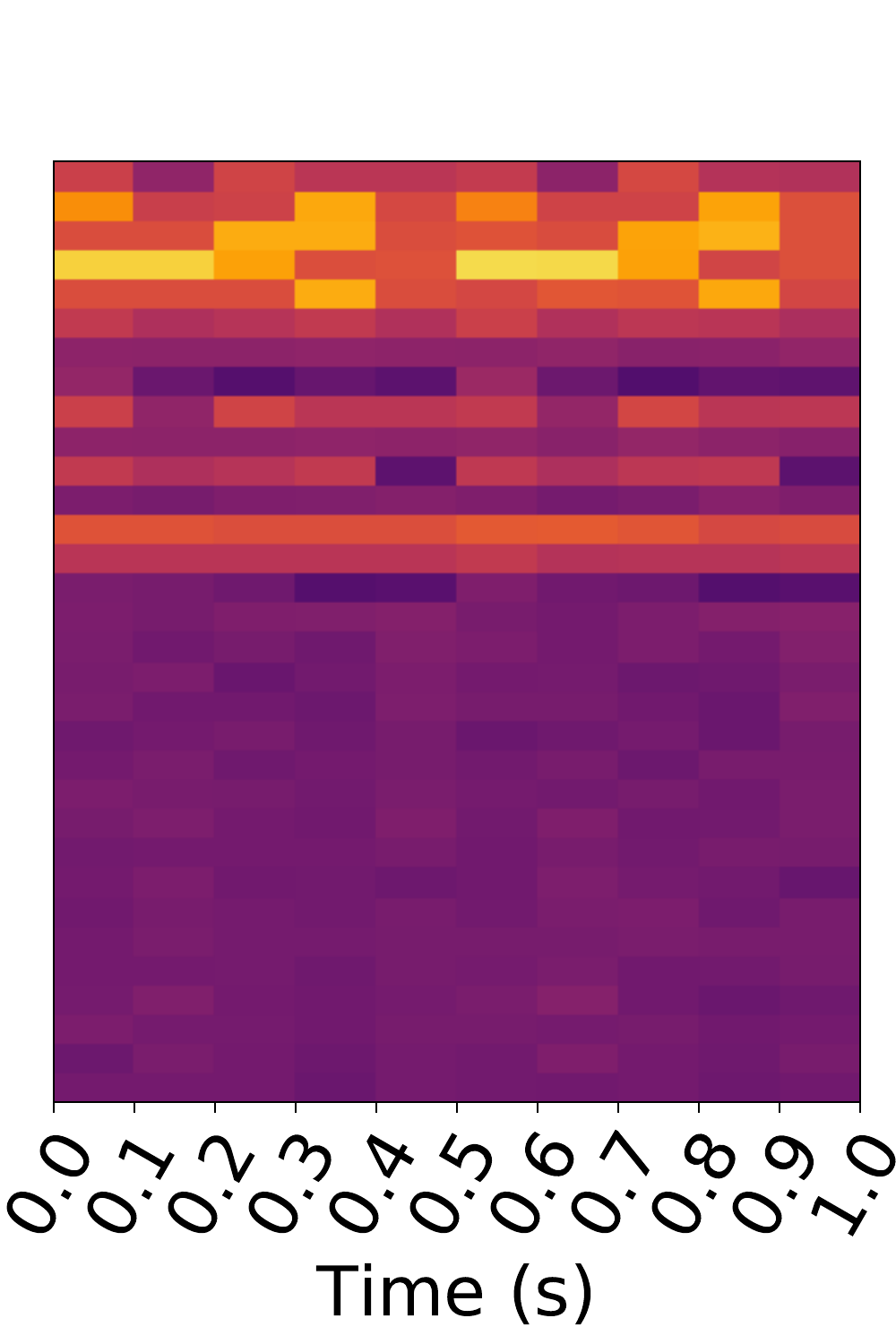}}
	& \hspace{-1.1in}\subfloat[CuttleSys]{\includegraphics[scale=0.24, viewport= -110 0 480 350]{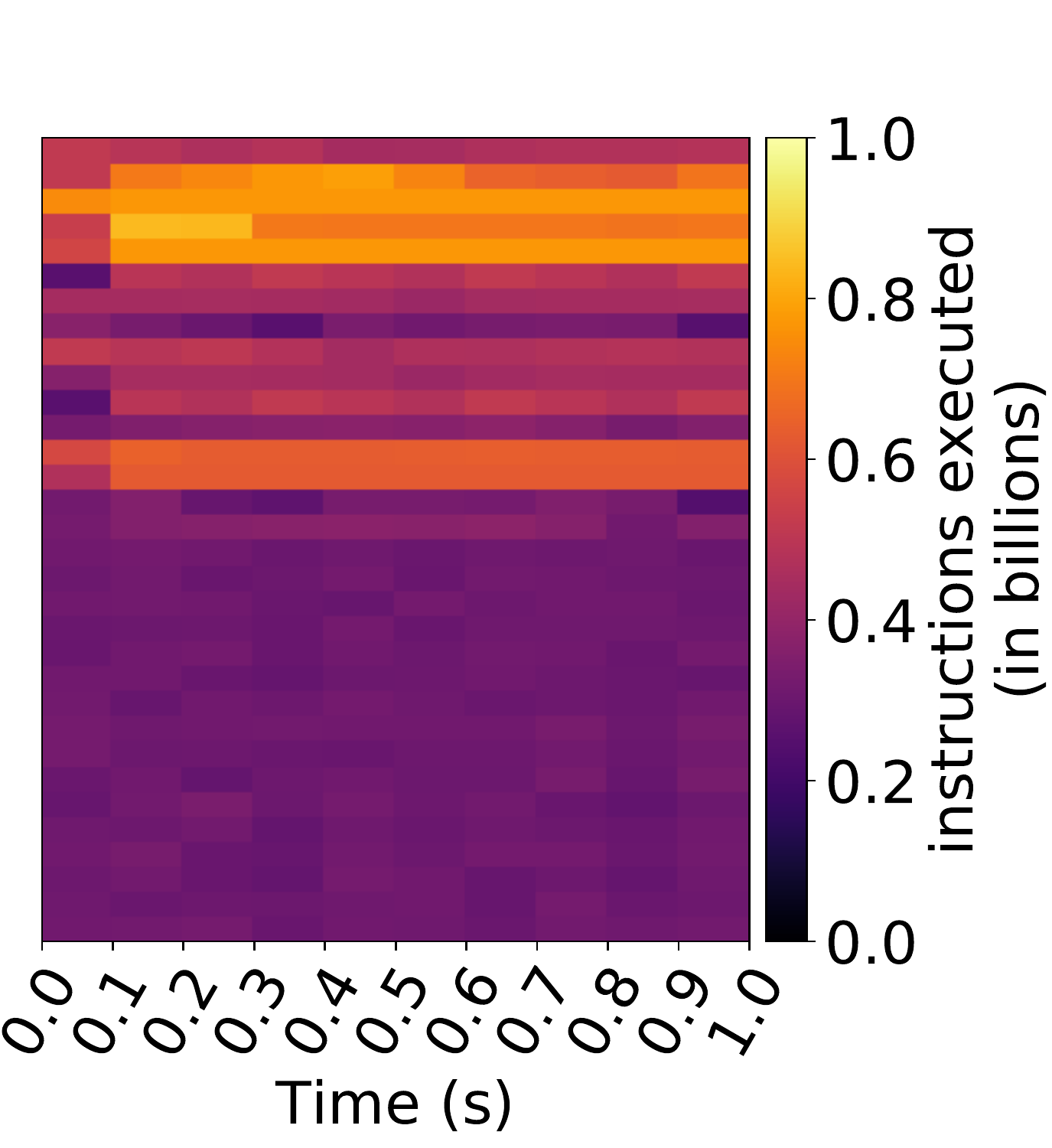}}\\
\end{tabular}
\caption{Instructions executed in each time slice (0.1s) on all cores with core-level gating, asymmetric cores, and CuttleSys. }
\label{fig:heatmap}
  \vspace{-0.12in}
\end{figure} 

\subsection{Core Gating and Asymmetric Multicores}
\label{sec:gating}

Fig.~\ref{fig:heatmap} shows the number of instructions executed on all cores 
in each timeslice over $1s$ with core-level gating and CuttleSys under a 70\% power cap. 
In the case of core-level gating, cores that consume the most power are turned off to meet the power budget 
and do not execute any instructions. In the case of asymmetric multicores, though all cores remain active, some jobs execute on small cores. 
We assume an unrealistic, oracle-like asymmetric multicore, where the number of big and small cores is determined to be the optimal, for 
a given workload, in each timeslice. 
To meet QoS, the latency-sensitive applications usually execute on big cores. 
For 70\% power cap, an additional 7 out of 16 batch applications execute on the big cores, while the remaining 9 applications execute on the small cores. 
CuttleSys also keeps all cores active, but portions of the cores might be turned off to meet the power budget. 

Fig.~\ref{fig:result}(c) quantitatively 
compares the total number of instructions executed by batch applications in 
(1) core-level gating without way-partitioning; (2) core-level gating with way-partitioning; 
(3) the oracle-like asymmetric multicore; and (4) CuttleSys, relative to no gating 
(all cores run in highest configuration) with no cache partitioning, for each power cap. 
QoS is satisfied for all Tailbench applications across all runs for core-level gating, oracle-like asymmetric multicore, and CuttleSys.
Results include all overheads of Sec.~\ref{sect:overheads}. 

For relaxed power caps (90\%), all cores can be turned on for the fixed-core multicores (core-level gating and asymmetric multicores), 
while parts of the cores need to be turned off with Cuttlesys, given the energy overhead of reconfiguration. 
Thus, CuttleSys performs worse than core-level gating and asymmetric multicores.

As the power caps decrease, however, CuttleSys outperforms core-level gating both without and with way-partitioning 
by 1.64$\times$ and 1.52$\times$ on average, and up to 2.65$\times$ and 2.46$\times$ respectively (Fig.~\ref{fig:result}(c)). 
CuttleSys also outperforms the oracle-like asymmetric multicore by 1.19$\times$ on average, and up to 1.55$\times$ for the most stringent power cap.
As power caps decrease, core-level gating turns off additional cores, 
while the oracle-like multicore executes more jobs on smaller cores.
The fine granularity of reconfigurable cores provides additional power/performance operating points, which permit better fine-tuning during 
power-constrained scenarios. These gains amortize the energy and scheduling overheads of CuttleSys. 

CuttleSys provides modest throughput gains over the oracle-like asymmetric multicore for relaxed power caps, 
as more batch jobs can execute on big cores in the asymmetric multicore. 
In real systems~\cite{armbiglittle}, the number of small and big cores is fixed. 
CuttleSys outperforms a typical multicore with 50\% big and 50\% small cores by 1.70$\times$, 1.65$\times$ and 1.50$\times$ at 90\%, 80\% and 70\% power caps respectively. 
The performance of this 50-50 multicore is the same as that of the oracle-like asymmetric system at 60\% and 50\% power cap, 
since all the batch applications run on small cores. 


\subsection{Dynamic behavior of CuttleSys}

We now show CuttleSys's behavior under varying load and power caps, 
and an example of core relocation. 

\vspace{-0.05in}
\subsubsection{Varying Load}

\begin{figure} 
\begin{tabular}{ccc}
  \hspace{-1.35in}\subfloat[]{\includegraphics[scale=0.38, viewport= -230 15 540 555]{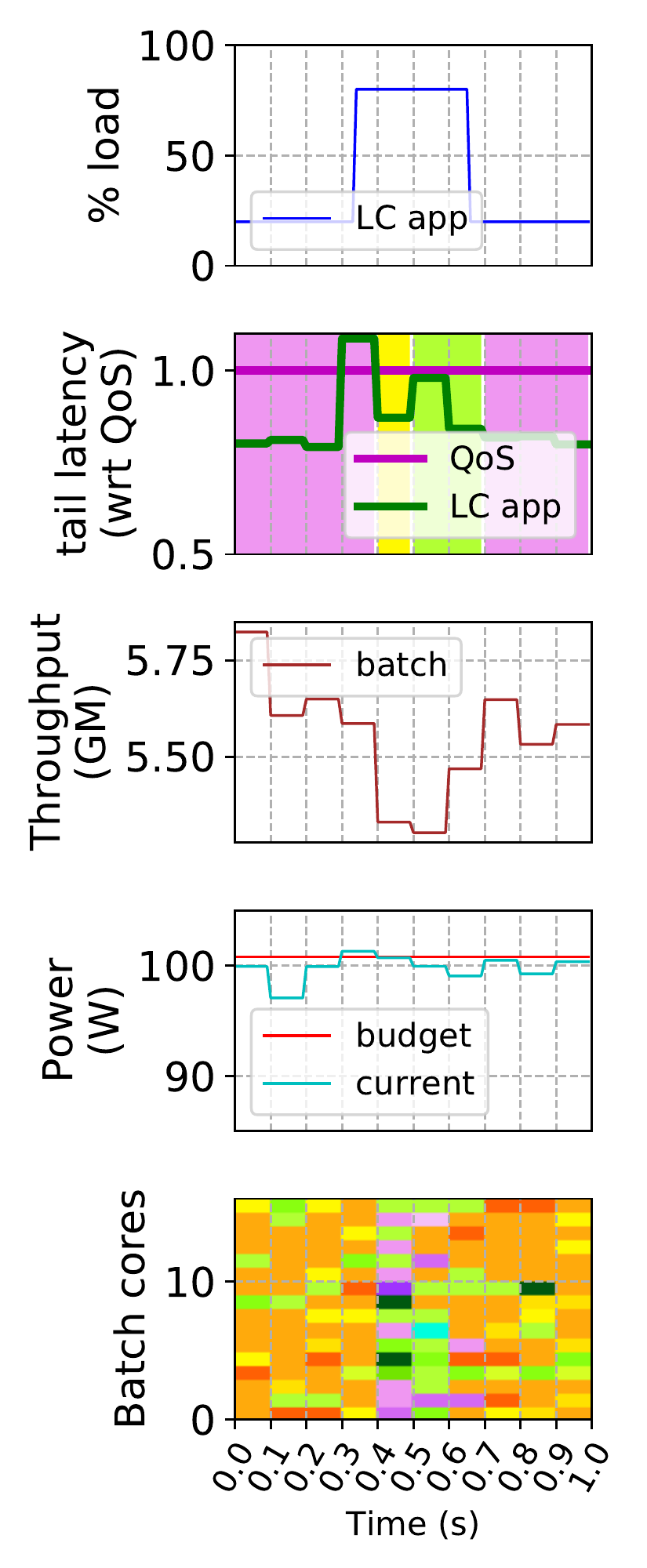}} & 
  \hspace{-3.4in}\subfloat[]{\includegraphics[scale=0.38, viewport= -298 15 540 555]{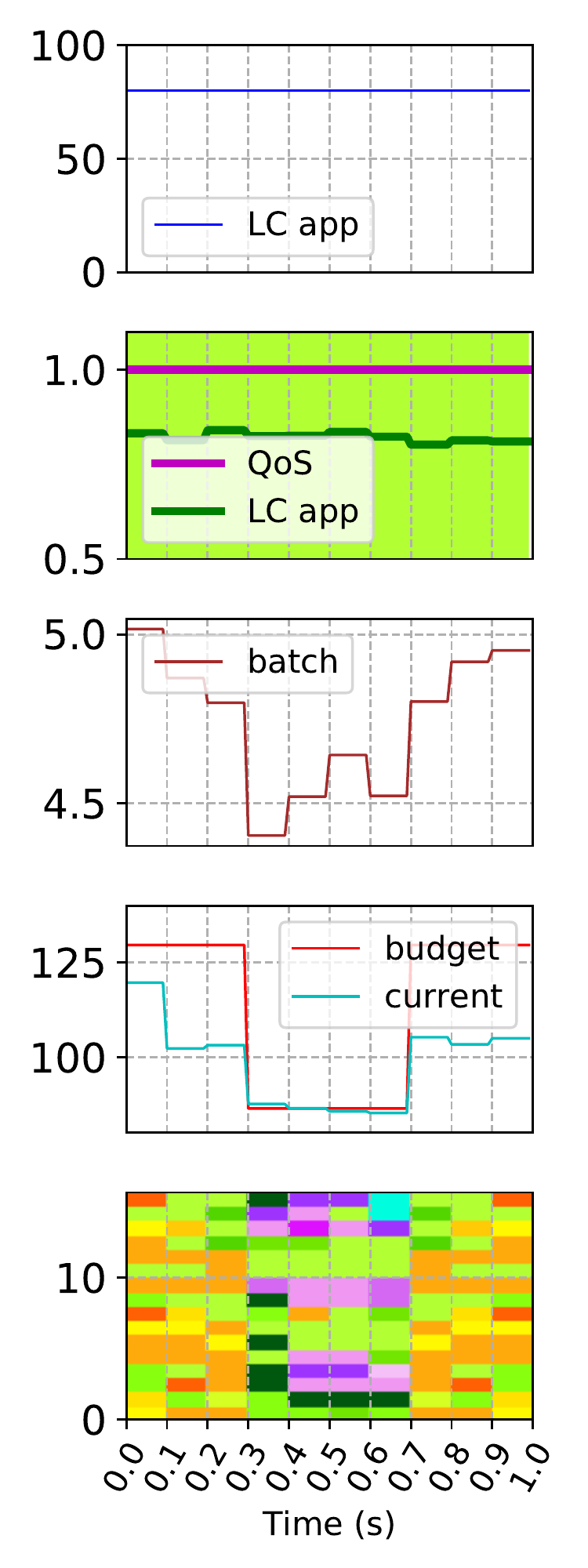}} & 
  \hspace{-3.6in}\subfloat[]{\includegraphics[scale=0.38, viewport= -313 15 540 555]{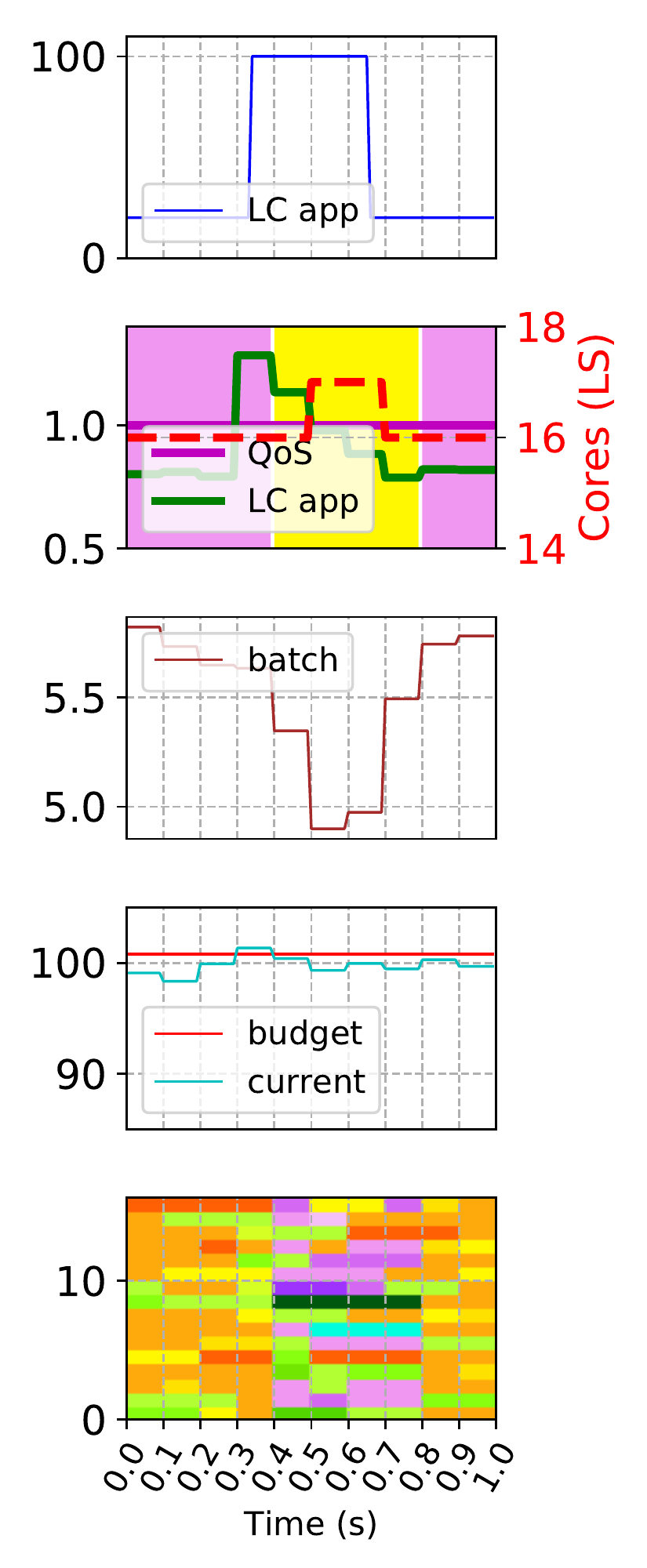}}\\[-0.08in]
  \multicolumn{3}{c}{\includegraphics[scale=0.5,viewport=85 0 500 70]{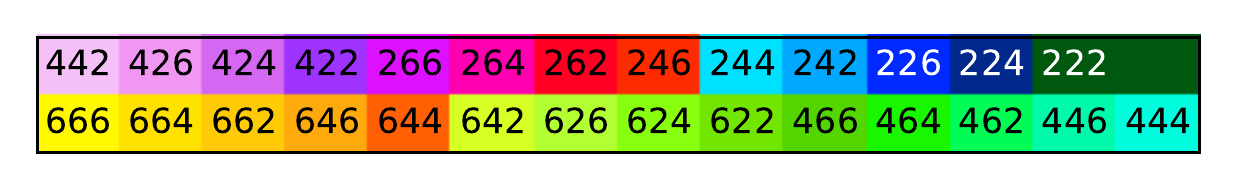}}
  \vspace{-0.1in}
\end{tabular}
\caption{CuttleSys under (a) varying input load, (b) varying power budget, and (c) example of core relocation. The table shows the colors corresponding to core configurations.}
\label{fig:varyload}
  \vspace{-0.08in}
\end{figure}

We vary the input load of the latency-critical application by simulating 
a diurnal pattern, while maintaining the power budget at 70\% of max. 
Fig.~\ref{fig:varyload}a shows the input load of the latency-critical application, 
its tail latency with respect to QoS, the throughput of batch applications, the total power consumed by the system, 
and the core configurations for batch applications 
for a colocation of Xapian with a mix of 16 
SPEC jobs. When load is low, cores running Xapian are configured to \{4,2,4\}, as shown by the background color.

As load increases, the tail latency also increases and violates QoS. 
Subsequently, CuttleSys configures the cores allocated to Xapian to the \{6,6,6\} configuration in the next time slice, 
after which QoS is met, and to \{6,2,6\} in the following time slice. 
Four cache ways are allocated to Xapian throughout the experiment.  
Under high load, Xapian consumes a significant fraction of the power budget, leaving less power 
for the SPEC applications. The cores running SPEC jobs therefore have to run 
in lower-performing configurations, and as a result achieve 
lower throughput. There is a brief interval in $t\in[0.3,0.4]s$ where the system violates its power budget. This is because 
the input load of Xapian increases in the middle of CuttleSys's decision interval, and the system needs to wait until the next interval before 
reconfiguring the cores. While this may briefly consume more power than required, it avoids ping-ponging between configurations 
due to short load spikes. 
When the load decreases, CuttleSys again reconfigures Xapian's cores to \{4,2,4\}, 
and set the remaining cores to higher configurations, thus increasing 
the throughput of SPEC jobs. 
 
\vspace{-0.05in}
\subsubsection{Varying Power Budget}

We now vary the power cap over time when running Xapian and a mix of SPEC applications, 
while maintaining a constant 80\% load for the latency-critical application. 
The power budget is set to 90\% and reduced to 60\% at $t=0.3s$. In this case (Fig.~\ref{fig:varyload}b), 
the cores running Xapian are configured to \{6,2,6\} and four cache ways for the entire duration of the experiment. 
When the power cap is reduced, Xapian still needs the same amount of power to meet its QoS, 
leaving a lower power budget for the SPEC workloads, which are configured 
to lower-performing configurations, decreasing their throughput. When the power cap is 
set back to 90\% at $t=0.7s$, the SPEC cores revert back to the higher 
configurations.

\vspace{-0.05in}
\subsubsection{Core Relocation}

Fig.~\ref{fig:varyload}c demonstrates an example co-scheduling Xapian with a mix of SPEC applications, 
where CuttleSys relocates cores to the latency-critical application to meet its QoS. 
As the load increases after $t=0.3s$, Xapian suffers a QoS violation, after which 
its allocated cores are reconfigured from \{4,2,4\} to the widest-issue configuration \{6,6,6\}. 
However, that is not sufficient to meet QoS in this case. 
Thus, CuttleSys reclaims a core from the batch applications, and assigns it to Xapian, at which point QoS is met. 
After the load drops back down to 20\%, tail latency also drops. Since now the latency slack is high enough (20\% unless otherwise specified), 
the extra core is yielded back to the batch applications. 
As a result of the core relocation, the SPEC applications time-multiplex on the reduced number of cores allocated to them, achieving lower throughput, which is recovered
when the core is returned. 

\vspace{-0.05in}
\subsection{Comparison with Flicker} 
\label{sec:flicker}

Flicker~\cite{flicker} is the most relevant prior work to CuttleSys. 
Flicker was proposed for multicore architectures 
running multi-programmed mixes of exclusively batch applications. It proposed 
3MM3 sampling~\cite{wu:3MM3} with RBF surrogate fitting~\cite{rbf1, rbf2, rbf3, rbf4, rbf5} to characterize the impact of 
core configurations, 
and a Genetic Algorithm (GA) for space exploration. 
Flicker relies on detailed per-configuration profiling, and is limited to core configurations, 
still allowing interference through the memory hierarchy. 
3MM3 requires sampling nine core configurations, 
which are then used by RBF surrogate fitting to get the complete 
performance and power profiles across all core configurations.
To get a meaningful sample for tail latency, the system needs to run for at least 10ms. 

We evaluated Flicker in two ways: 
a) we set the profiling period to 10ms and profile the applications for a total of 90ms, 
search the best configuration that meets the QoS and power budget and maximizes 
the throughput using GA (takes 2ms), and run the system in that configuration for the 
remaining 8ms; b) Flicker only manages batch applications, and we set the cores assigned to latency-critical jobs
to the highest -- \{6,6,6\} -- configuration, which reduces the power budget 
available for batch jobs. In this case, since we only predict 
throughput and power, we can directly apply the 3MM3 and RBF techniques over $1ms$ samples. 
Overall, we profile for $9ms$, and run GA for $2ms$. In both cases, 
we have to run the latency-critical service in lower configurations 
for extended periods of time. Since QoS is defined with respect to the 99$^{th}$ percentile latency, 
even $1ms$ of slow requests is enough to violate QoS. 
As a result, we see extensive QoS violations by over an order of magnitude 
for the first methodology, and by 1.5$\times$ for the second. 

\begin{figure}
	\begin{tabular}{cc}
	\centering
	\includegraphics[scale=0.29,bb=0 0 30 30,viewport=0 25 370 390]{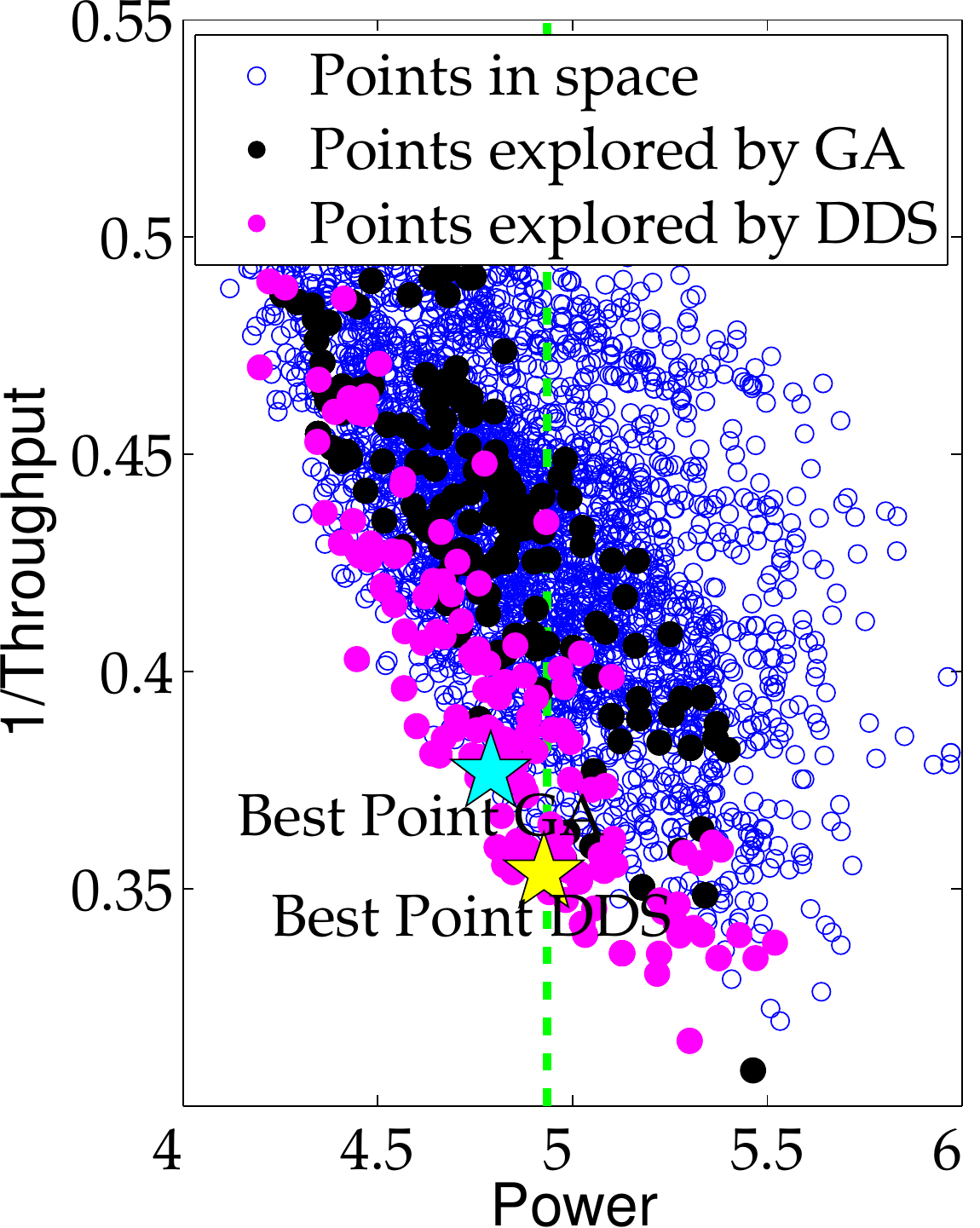} & 
	\includegraphics[scale=0.39,bb=0 0 30 30,viewport = 30 40 400 330]{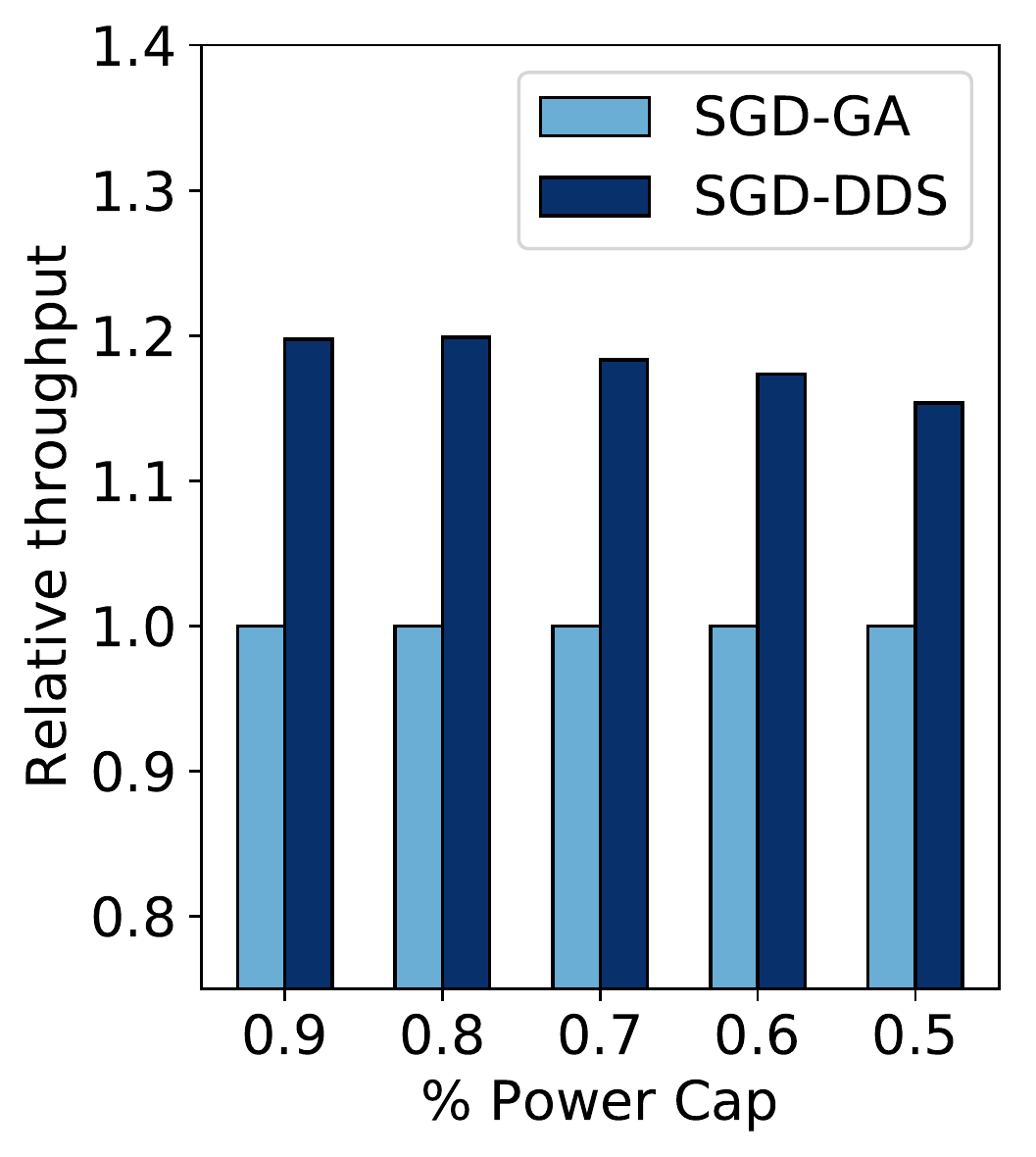}
\end{tabular}
	\caption{(a) Comparison of DDS vs GA's ability to explore the design space. 
(b) Throughput with DDS and GA under different power caps using SGD for inference. }
	\label{fig:vsflicker_dds_ga}
\vspace{-0.08in}
\end{figure}

\begin{wrapfigure}[11]{r}{0.204\textwidth}
\centering
\includegraphics[scale=0.27,bb=0 0 30 30,viewport= 450 20 50 260]{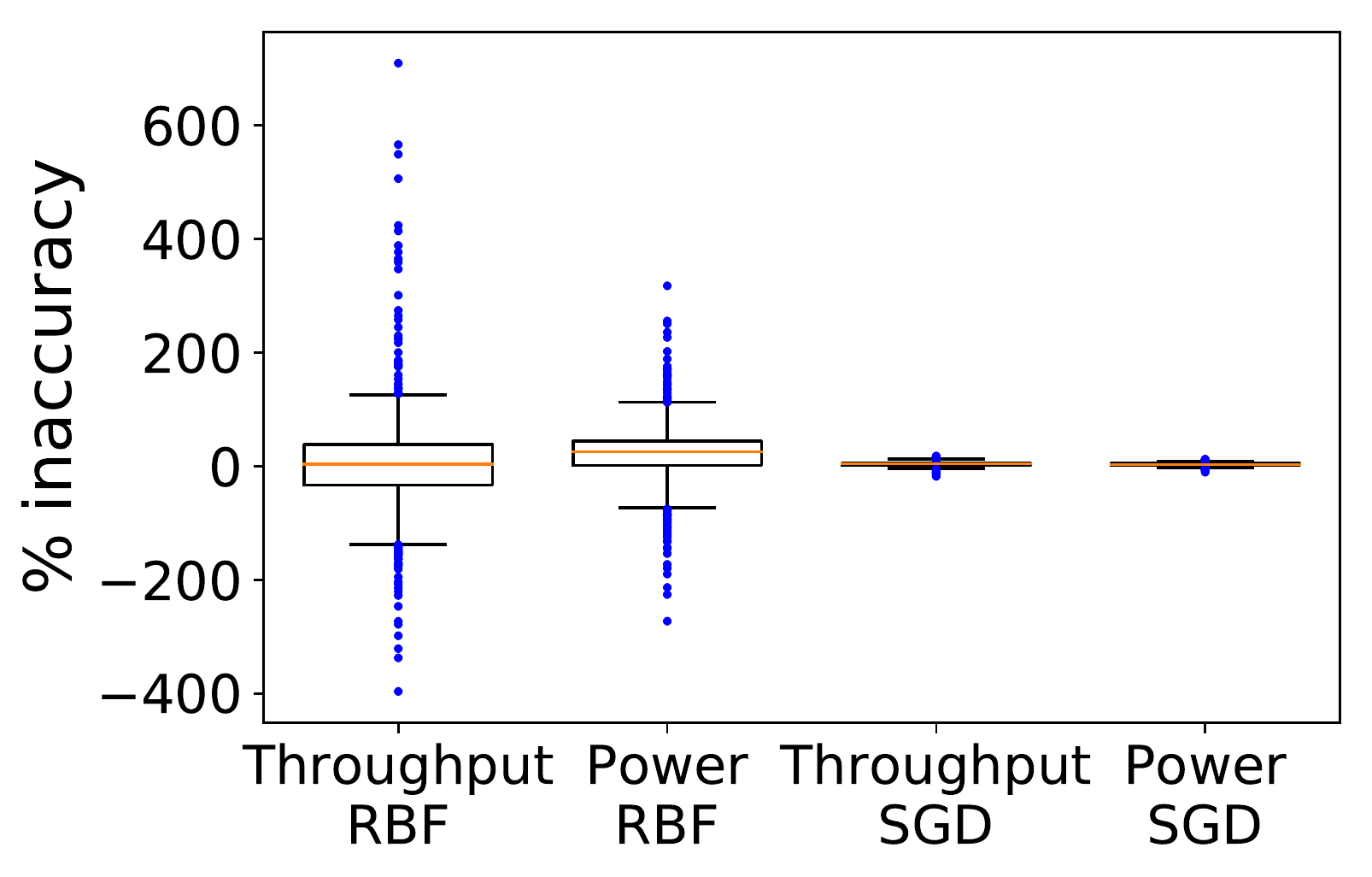}
\vspace{-0.08in}
\caption{Predicted errors in performance (1) \& power (2) with RBF.}
\label{fig:flicker-error}
\vspace{-0.08in}
\end{wrapfigure}

We now compare the individual techniques used in Flicker and CuttleSys. 
Flicker requires 9 samples for characterization, while SGD only uses 2 samples. 
To show a fair comparison for the characterization mechanisms, we show the prediction error of the RBF-based approach 
in performance and power in Fig.~\ref{fig:flicker-error} 
when using 3 samples from the full 100ms timeslice (the algorithm was unable to converge when using two samples). 
The error is dramatically higher for Flicker with 3 samples, with outliers reaching up to 600\%. 
Thus, with the same amount of information, the SGD-based reconstruction clearly outperforms the RBF-based approach.  

Next, we compare the exploration algorithms, GA and DDS. Fig.~\ref{fig:vsflicker_dds_ga}a shows 
a subset of points in the entire space, as well as the points explored by both DDS and GA. 
The black dots represent the points explored by GA, while the pink dots represent the points 
explored by DDS. We can see that DDS explores more points on the pareto-optimal front and thus, 
obtains a higher-quality configuration with better throughput compared to GA, 
shown by blue and yellow stars respectively, under a given power budget, shown by the dotted green line. 

To quantitatively compare DDS with GA, we applied GA during the optimization phase instead of DDS, and used SGD for reconstruction 
in our 32-core system. Fig.~\ref{fig:vsflicker_dds_ga}b shows the comparison of the geometric mean of throughput 
of CuttleSys with SGD and GA across different power caps.
Using DDS for optimization offers a performance improvement of up to 19\% compared to GA for a 32-core system. 
This can be attributed to the fact that the GA algorithm is relatively slow in exploring a highly-dimensional search space compared to DDS.
Also, the optimization algorithm is required to explore a higher number of configurations $27*4 = 108$ (including LLC allocations), compared to only $27$ core configurations in~\cite{flicker}. 
We also note that, the performance improvement is higher at lower power caps, 
as a large subset of configurations does not violate the power budget, and DDS can quickly search through the large space. 
As the power constraints become more stringent, fewer configurations are valid, enabling GA to find
the best configurations in a given amount of time. The improvement is the smallest 
for a 50\% power cap as at that point, all cores often have to operate in their lowest configurations, 
and may even need to be switched off to meet the power budget. 

\vspace{-0.12in}

\section{Conclusions}
\label{conclusions}


We present CuttleSys, an online and practical resource management system for reconfigurable multicores, 
which quickly infers the performance and power consumption of each co-scheduled application across all 
core configurations and cache allocations, and arrives at a suitable configuration that meets QoS for the 
latency-critical services, and maximizes throughput for the batch workloads, under a power budget. 

We evaluate CuttleSys across a set of diverse latency-critical and batch workloads, 
and showed that the system meets both the QoS and power budget at all times, 
while achieving significantly higher throughput for the batch applications than 
previous work, including core-level gating and Flicker. We also quantified the 
inference errors of the reconstruction algorithm in CuttleSys, and showed that 
they are low in all cases. 




\bibliographystyle{IEEEtranS}
\bibliography{ref}

\end{document}